\documentclass[prm,preprint,showpacs,amsmath,amssymb,aps,unsortedaddress]{revtex4-1} %,preprintnumbers,superscriptaddress,unsortedaddress
\usepackage{graphicx}% Include figure files
\usepackage{dcolumn} % Align table columns on decimal point
\usepackage[xdvi]{epsfig}
\usepackage[usenames,dvipsnames]{color}
\usepackage{txfonts}
\usepackage{subfigure}
\usepackage[utf8x]{inputenc} %utf8x utf8
\usepackage[T1]{fontenc}
\usepackage{amsfonts}

  \newcommand{\hevo}[2]{{#1} } % version without markups
\begin{document}
\title{Dynamical tilting in halide perovskites}

\author{Donat J. Adams}
\affiliation{University of Bern, Bern, Switzerland}

\email{donat.adams@unibe.ch}
 \author{Sergey V. Churakov}

\affiliation{University of Bern, Bern, Switzerland; Laboratory for Waste management, Paul Scherrer Institute, Villigen-PSI, Switzerland}
\date{\today}

%\pacs{64.10.+h, 63.20.Ry, 71.15.Mb}

\begin{abstract}
Naturals and synthetic perovskites are widely used functional materials thanks to their particular physical properties, such as superconductivity, ferroelectricity and photo-activity. Many of these properties are related to static or dynamic motion of octahedral units; Yet a full understanding of the relationships between perovskite crystal structure, chemical bonding and physical properties is currently missing. Several studies indicate the existence of dynamic disorder generated by anharmonic motion of octahedral units e.g. in halide perovskite structures. In this work, we derive all possible space groups for simple perovskites $\rm ABX_3$ with dynamical octahedral tilting. The derived space groups  extend the well established space group tables for static tiltings by [Glazer, A. Acta Cryst. B 1972, \textbf{28}, 3384-3392; Aleksandrov, K. Ferroelectrics 1976, \textbf{24}, 801-805; Howard, C.J., Stokes, H.T. Acta Cryst. B 1998, \textbf{54}, 782-789]. We demonstrate ubiquity of dynamical tilting by analysing the structural data for perovskites reported in recent scientific publications and discuss the signature of dynamic tilting in the corresponding structures. Finally, we discuss the possible influence of dynamic disorder on the physical properties of halide perovskites.
\end{abstract}

\maketitle

\section{Introduction} %
%\subsection{What is known}  In Crystals wird einiges \"uber Perovskite Publiziert, z.B \"uber SrTio3 (https://www.mdpi.com/2073-4352/9/11/580) aber vor allem \"uber Halid-Perovskite \cite{li2018}. Dort könnte man ankn\"upfen (siehe weitere Punkte).

Solids with perovskites structure are one of the most widely used functional materials in the microelectronic, thanks to their exceptional physical properties, such as superconductivity \cite{bednorz1986}, ferroelectricity \cite{fan2015} and photo-activity \cite{granados2020}. The origin and underlying physical mechanism of these properties are intriguing from the theoretical and the application point of view. Yet a full understanding of the relationships between perovskite crystal structure, chemical bonding and physical properties is currently missing. Some observations appear counterintuitive at the first glance. For example, structural defects usually reduce the yield of photoactive materials because the presence of defects decreases the carrier density and open up non-radiative recombination paths for the photo-induced generation of charge carriers. $\rm CH_3NH_3PbI_3$ is known to have an intrinsically disordered perovskite type structure with large number of defects and dynamic disorder, which are very unusual for highly efficient photovoltaic structures. It is therefore intriguing to understand how disordered solution-processed materials such as halide perovskites as e.g. $\rm CH_3NH_3PbI_3$  with $X=$I, Br, or Cl \cite{li2018, bechtel2019} can possess such a high photoactive yield efficiency which can even compete with high quality crystalline semiconductors. \\
Two types of disorder are known for perovskite based structures: mostly non polar rotations of the octahedra \cite{zhao2016}, and polar displacements \cite{wright2016}. Interestingly both types of structural disorder can potentially interact with electrons, either through the acoustic deformation potential scattering \cite{zhao2016, neukirch2016} or through Frohlich type polar interactions \cite{wright2016}. The structural disorder is thus expected to influence the carrier mobility \cite{karakus2015} in metal halide perovskites. Furthermore it has been suggested, that ionic displacement could be the source of giant dielectric constants \cite{samara1990} in lead halide perovskites and polaronic conductors.

It has yet been overlooked so far, that atomic displacements and the corresponding disorder can be inherent to some perovskite phases within certain cell geometries, ions are confined by a potential energy surface with multiple local energy minima (see e.g.~Fig.~\ref{fig:states_single_well_double_well}) and separated by small energy barriers (40~meV, for $\rm CH_3NH_3PbI_3$  \cite{beecher2016}) . If these minima are close in energy, the transition between these minima is enhanced and can be excited by small perturbations \cite{adams2016}.
%Polarity typically affects optical and electrical properties, but it is not indisputably proven for $\rm CH_3NH_3PbI_3$ \cite[e.g.][]{beilsten2015}.

% relativley short lived and therefore call
In this article, we focus on the role of mostly non polar rotations of the octahedra, which are ubiquitous in halide perovskites. They have been observed for example in $\rm CsPbCl_3$ \cite{fujii1974}, $\rm CH_3NH_3PbBr_3$ and $\rm CH_3NH_3PbCl_3$ \cite{chi2005, swainson2003, swainson2015}. This kind of disorder is dynamic, i.e. displacements are created and annihilated permanently, whiles as the disorder persists. Related to the observations in the cubic phase of  $\rm CH_3NH_3PbI_3$ through inelastic X-ray scattering the term `dynamic disorder' has been coined \cite{poglitsch1987, egger2018} (see also the review of Whalley et al. \cite{whalley2017}). In this article we will refer to these octahedral rotations as `dynamic tilting'. The presence of dynamic tilting e.g in halide perovskites is well documented \cite{beecher2016}. The dynamic instabilities are involved in the series of phase transitions in $\rm CH_3NH_3PbI_3$ from the orthorhombic phase (\emph{Pbn}2$_1$ space group), to tetragonal structure (\emph{I}4\emph{cm} space group) at around 160~K, and finally at approximately 330~K to the pseudocubic tetragonal phase (\emph{P}4\emph{mm} space group) \cite{onoda1992}.

On the other hand, when it comes to structure determination it is surprising that the dynamic instability has been given little consideration yet, although many perovskites show dynamic instability especially in the high temperature phase.

\hevo{The mechanism stabilising the dynamical tilts is explained as follows: The energies of the oscillation modes in the doublewell potential are usually denser than those in a single well. Furthermore, dynamic tilting is favoured when the ground state of the double potential is (nearly) degenerate, leading to a \emph{particularly high density of states at the lowest energy}. Generally, the high temperature phase is higher in potential energy than the low temperature phase. This energy difference is outweighed by the free energy, which is the relevant thermodynamic potential at temperature.}{A2}

\hevo{For halide perovskites, this model predicts a generally high sensitivity to external influences such as electric fields, temperature, humidity, or mechanical stress \cite{lin2021, ugur2020}. We relate it to the observed instability on the surface of the potential energy. Interestingly it is precisely these instabilities and the resulting dynamic tilting that can lead to high mobility of charge carriers and in turn might be the premise of the high efficiency of perovskite solar cells. }{A3}

We therefore systematically explore possible dynamic rotations of the octahedra in perovskite structures. Then we explore the most important experimentally accessible signatures of dynamic tilting. Our hypothesis is that many perovskites exhibit dynamic tiling, especially in the group of halide perovskits. Accordingly, we review most recent publications reporting dynamic tilting, in order to substantiate our hypothesis. Finally, we will discuss the possible implications of dynamic tilting for the physical properties of for halide perovskites such as $\rm CH_3NH_3PbI_3$ considering the electronic structure, vibrational properties especially phonon dispersion, the electron-phonon coupling and efficiency in photovoltaics.

\section{Method}
%% structure
% determination of space groups
% consequences of the dynamic tiling e.g. negative thermal expansion

To derive possible tilting systems, a 2$\times$2$\times$2 supercell of the ideal cubic ABO$_3$ perovskite containing 40 atoms was setup in which the rigid octahedra were tilted in three spatial directions. The Glazer notation \cite{glazer1972} has been used to denote by $a^+$ an octahedral tilting by angle $a$ where two succeeding tilting are in phase and $ a^-$ for succeeding octahedra tilted out of phase. These tilts lead to a contraction of the two lattice parameters which are perpendicular to the tilting axis --- see also Fig.~\ref{fig:undist_octahedra}. This reduction has been implemented using three dimensional rotation matrices. They preserve the bonding distance within the octahedron without further (collective) constraints on the structure.

\begin{figure}  % Requires \usepackage{graphicx}
 \centering
 \includegraphics[width=0.9\textwidth]{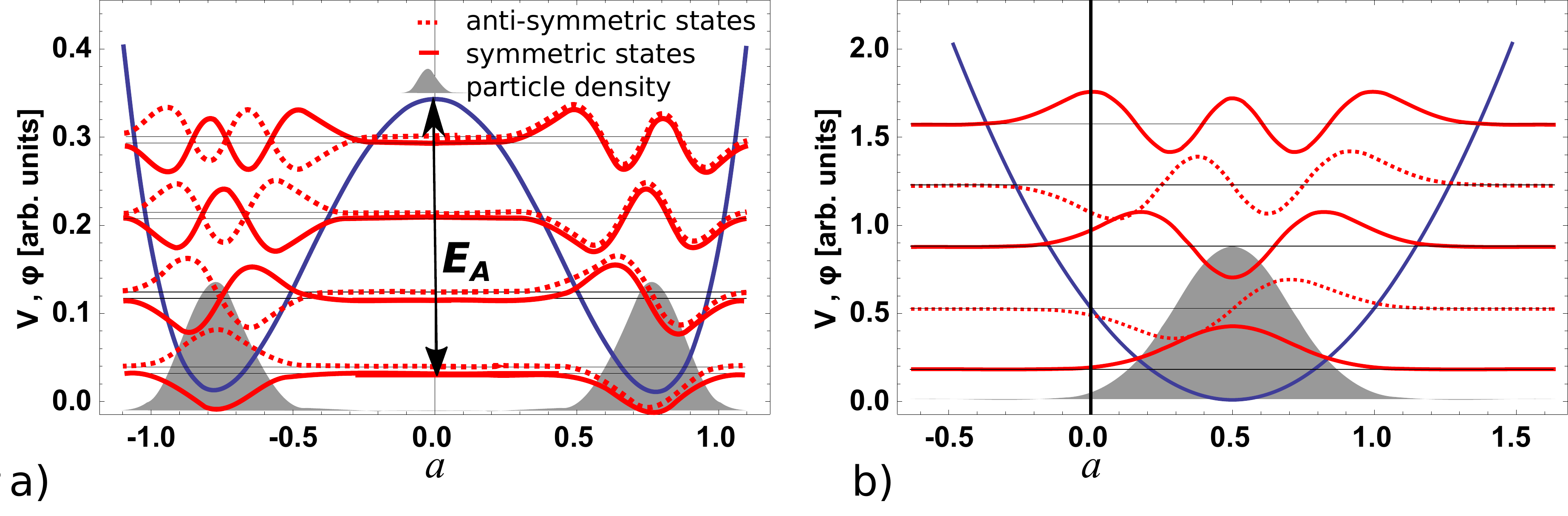}\\ %
 \caption{Sketch of the potential energy surface as a function of the rotational angle $a$ of the BX$_3$ octahedron. The atomic configurations generating these potentials are shown in Fig.~\ref{fig:undist_octahedra}. \textbf{a)} Dynamic tilting: the rotational degree of freedom is subject to a double well potential $V(a)$. The resulting \emph{ionic wave} function $\varphi(a)$ with lowest energy shows two distinct localisations in both potential minima and so does the particle density $|\varphi(a)|^2$ (shaded in grey). However, the average tilting angle is at $a=0$ --- i.e. in between the two minima.\textbf{ b)} Static tilting: The octahedral rotational degree of freedom is subject to a single potential with a minimum at $a\neq 0$. The resulting \emph{ionic wave} function $\varphi(a)$ with lowest energy shows a single localisation at the potential energy minimum. }\label{fig:states_single_well_double_well}
\end{figure}

\begin{figure}%[H]  % Requires \usepackage{graphicx}
 \centering
 \includegraphics[width=0.9\textwidth]{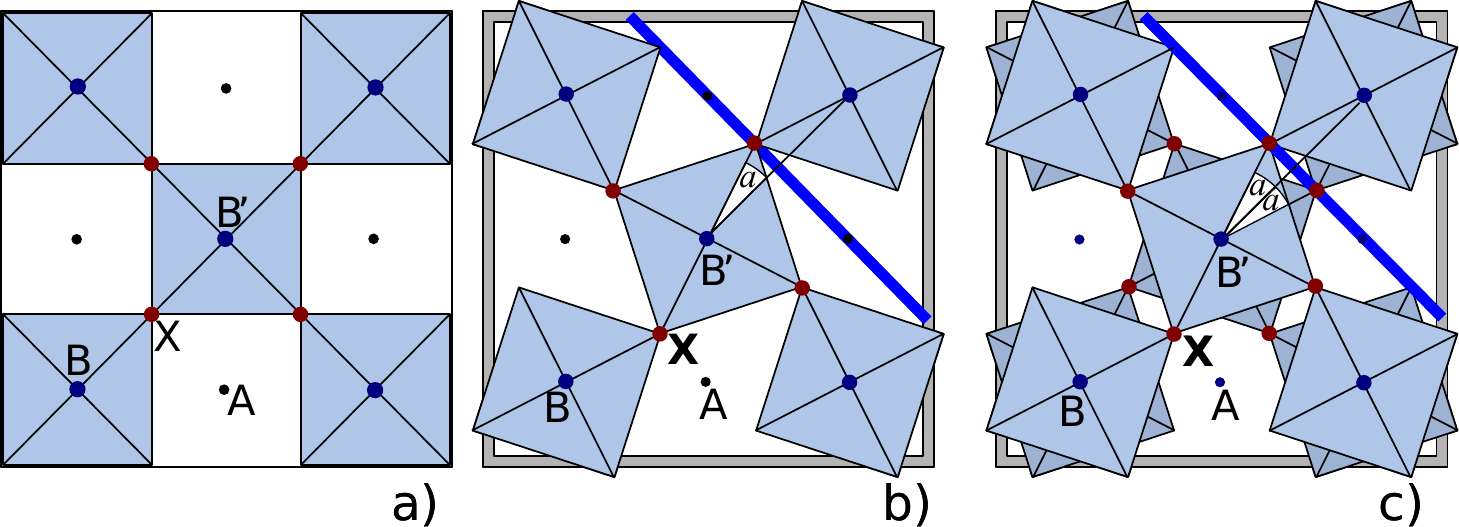}\\
 \caption{a) Ideal perovskite with untilted octahedra. b) Octahedral tilting by angle $a$ allows to reduce the distance between the cations B and B' while the bond length between B and X is maintained. The result is a reduction of the cell volume (highlighted in grey) due to less volume assigned to the A cations. c) Along the tilting axis two subsequent octahedra can be tilted in antiphase (here, notation $a^-$) or in phase (notation $a^+$, see subfigure b). Cross sections of the potential energy surface (PES) along the blue lines corresponding to octahedral rotations are given in Fig.~\ref{fig:states_single_well_double_well}.}\label{fig:undist_octahedra}
\end{figure}

As explained in the introduction, recent studies have lead to the discovery of dynamic tilts \cite{marronnier2017, adams2016}.  They result from octahedra confined in the structure by potential energy surface (PES) with multiple energy minima  --- see Fig.~\ref{fig:states_single_well_double_well}. In this study, we only consider systems with symmetric arrangements of minima of the PES related by a mirror plane. In these structures the tilt-angle oscillates between positive and negative amplitude spending most of the time at $|a|>0$ whereas the average position is still $a=0$ ---  see Fig.~\ref{fig:states_single_well_double_well}. However, due to the non-zero average tilt-amplitude time averages of cell parameters perpendicular to the tilting axis become reduced compared to the untilted cubic structure. We denote these tilting by $a^d$ and consider them by leaving the atomic positions unchanged and reducing the cell parameters perpendicular to the tilting axis. \\

\subsection{Classification of dynamic tilt systems}
The possible combinations of dynamic tilts and their combinations with static tilts were applied and the resulting structures were classified using the crystallographic tool FINDSYM \cite{stokes2005}. It determines the primitive unit cell, lattice vectors, point group of the lattice and thus the unique space-group of crystal structure. \\
For static tilts Howard and Stokes \cite{howard1998} showed, that the 25 distinct isotropy subgroups of static tilts can be reduced to 15. They consider only `simple' tilt systems, in which the tilts around a particular axis have the same magnitude and either the same sign (the + pattern) or alternating sign (the - pattern). This results in the exclusion of tilt systems, that accidently show the same tilt angle but different type of tilting ($+,-$) on different axis. \\
Application of this concept to the dynamic tilts i.e. excluding tilting systems showing the same tilting angle but different type of tilting along different axis 19 different tilting systems have been derived. Their space groups are reported in Tab.~\ref{tab:classification}.

\begin{table}
 \centering
\begin{tabular}{ l|c  c|c c c }       % after \\: \hline or \cline{col1-col2} \cline{col3-col4} ...
 Tilt system& Space group & No & $\mathbf{a}$& $\mathbf{b}$& $\mathbf{c}$\\   \hline
%$ a^+ b^+ c^+ $&\\ % 1
%$ a^+ a^+ b^+ $&\\ % 2
%$ a^+ a^+ a^+ $&\\ % 3
%$ a^+ b^+ c^- $&\\ % 4
%$ a^+ a^+ b^- $&\\ % 5
%$ a^- a^+ b^+ $&\\ % 6
%$ a^+ a^+ a^- $&\\ % 7
%$ a^+ b^- c^- $&\\ % 8
%$ a^- a^+ b^- $&\\ % 9
%$ a^- a^- b^+ $&\\ % 10
%$ a^- a^- a^+ $&\\ % 11
%$ a^- b^- c^- $&\\ % 12
%$ a^- a^- b^- $&\\ % 13
%$ a^- a^- a^- $&\\ % 14
%$a^0 a^+ b^+ $&\\ % 15
%$a^0 a^+ a^+ $&\\ % 16
%$a^0 a^- b^+ $&\\ % 17
%$a^0 a^- a^+ $&\\ % 18
%$a^0 a^- b^- $&\\ % 19
%$a^0 a^- a^- $&\\ % 20
%$ a^+a^0a^0 $&\\ % 21
%$ a^-a^0a^0 $&\\ % 22
%$a^0a^0a^0 $&\\ % 23
 $ a^- b^- c^d $ & \emph{C}2/\emph{m} & 12 &$2a_p$&  $-2c_p$& -$a_p+b_p$\\ % 24
 $ a^+ b^d c^+ $ & \emph{Immm} & 71 &$2a_p$&  $2b_p$&  $2c_p$ \\ % 25
 $ a^d b^+ c^- $ & \emph{Cmcm} & 63 &-$2a_p$&  $2c_p$&  $2b_p$ \\ % 26
 $ a^- b^d c^d $ & \emph{Fmmm} & 69 & $2a_p$&  $2b_p$&  $2c_p$ \\ % 27
 $ a^d b^+ c^d $ & \emph{Cmcm} & 65 &-$2a_p$&  $2c_p$&  $b_p$ \\ % 28
$ a^d b^- c^d $ & \emph{Fmmm} & 69 & $2a_p$&  $2b_p$&  $2c_p$ \\ % 29
$ a^+ b^d c^d $ & \emph{Cmmm} & 65 & $2c_p$&  $-2b_p$&  $a_p$ \\ % 30
%$ a^d b^d c^- $ & \emph{Fmmm} & 69 & \\ % 31 = 29
% $ a^d b^d c^+ $ & \emph{Cmmm} & 65 & \\ % 32 = 28
$ a^d b^d c^d $ & \emph{Pmmm} & 47 & $c_p$&  $b_p$& -$a_p$ \\ % 33
 $ a^- a^- b^d $ & \emph{Imma} & 74 & $a_p-b_p$&  $-2c_p$&  $a_p+b_p$ \\ % 34
% R3020$ a^- a^+ b^d $ & \emph{Cmcm} & 63 & $-2c_p$&  $2a_p$&  $-2b_p$ \\ % 35
% R3020 $ a^- a^d b^- $ & \emph{C}2/\emph{m} & 12 &-$2a_p$&  $-2b_p$&  $a_p+c_p$\\ % 36
% R3020 $ a^- a^d b^+ $ & \emph{Cmcm} & 63 & $2b_p$& -$2a_p$&  $2c_p$\\ % 37 =
 $ a^+ a^+ b^d $ & \emph{I}4/\emph{mmm} & 139 & $2a_p$&  $2b_p$&  $2c_p$\\ % 38
% R3020$ a^+ a^d b^- $ & \emph{Cmcm} & 63 & $-2b_p$&  $2c_p$& -$2a_p$\\ % 39
%\hline
% R3020 $ a^+ a^d b^+ $ & \emph{Immm} & 71 & $2a_p$&  $2b_p$&  $2c_p$ \\ % 40
% R3020$ a^+ a^d b^d $ & \emph{Cmmm} & 65 & $2c_p$&  $-2b_p$&  $a_p$\\ % 41
% R3020 $ a^- a^d b^d $ & \emph{Fmmm} & 69 & $2a_p$&  $2b_p$&  $2c_p$\\ % 42
$ a^d a^d b^- $ & \emph{I}4/\emph{mcm} & 140 &$-a_p-b_p$&  $a_p-b_p$&  $2c_p$\\ % 43
$ a^d a^d b^+ $ & \emph{P}4/\emph{mbm} & 127 & $a_p+b_p$& -$a_p+b_p$&  $c_p$\\ % 44
$ a^d a^d b^d $ & \emph{P}4/\emph{mmm} & 123 & $b_p$&  $a_p$&  $-c_p$ \\ % 45
% R3020$ a^- a^+ a^d $ & \emph{Cmcm} & 63 & $-2c_p$&  $2a_p$&  $-2b_p$ \\ % 46
% R3020$ a^- a^- a^d $ & \emph{Imma} & 74 & $-2c_p$& -$a_p+b_p$& $a_p+b_p$ \\ % 47
% R3020$ a^+ a^+ a^d $ & \emph{I}4/\emph{mmm} & 139 &  $2a_p$&  $2b_p$&  $2c_p$\\ % 48
% R3020 $ a^- a^d a^d $ & \emph{I}4/\emph{mcm} & 140 & $-b_p-c_p$& $-b_p+c_p$& -$2a_p$ \\ % 49
% R3020$ a^+ a^d a^d $ & \emph{P}4/\emph{mbm} & 127 & $b_p+c_p$& $-b_p+c_p$& $a_p$\\ % 50
$ a^d a^d a^d $ & \emph{P}$m\overline{3}m$ & 221 &  $-b_p$& $a_p$& $c_p$\\ % 51
 $a^0 a^- b^d $ & \emph{Fmmm} & 69 &   $2a_p$& $2b_p$& $2c_p$\\ % 52
 $a^0 a^+ b^d $ & \emph{Cmmm} & 65 &  -$2a_p$& $2c_p$& $b_p$\\ % 53
% $a^0 a^d b^- $ & Fmmm & 69 &  $2a_p$& $2b_p$& $2c_p$\\ % 54 = 52
%$a^0 a^d b^+ $ & \\ % 55 = 53
$a^0 a^d b^d $ & \emph{Pmmm} & 47 &  $c_p$& $b_p$& -$a_p$\\ % 56
% R3020 $a^0 a^- a^d $ & \emph{Fmmm} & 69 &  $2a_p$& $2b_p$& $2c_p$\\ % 57
% R3020$a^0 a^+ a^d $ & \emph{Cmmm} & 65 &-$2a_p$& $2c_p$& $b_p$ \\ % 58
$a^0 a^d a^d $ & \emph{P}4/\emph{mmm} & 123 & $c_p$& $b_p$& -$a_p$\\ % 59
$ a^da^0a^0 $ & \emph{P}4/\emph{mmm} & 123 & $c_p$& $b_p$& -$a_p$ \\ % 60
%       \hline
\end{tabular}\\
 \caption{19 perovskite-type structure tilt systems with dynamic tilting. The notation of Glazer \cite{glazer1972} was extended by a dynamic tilt $a^d$, indicating that the corresponding octahedra oscillate between two positions (positive and negative amplitude). This tilt can be observed instantaneously and locally while as the octahedra appear untilted when averaged over time or/and space.
The unit cell is given in terms of the pseudo-cubic axes $a_p$, $b_p$ and $c_p$, which are parallel to the axes of the tilting by angle $a$, $b$ or $c$, respectively. The distortion of the pseudo-cubic axes is given by $a_p=2 d\cdot \cos(b) cos (c),\; b_p =2 d\cdot cos(a) cos(c),\; c_p=2d\cdot \cos(a)cos(b)$, where $d$ corresponds to the B-X bond distance.
}\label{tab:classification}
\end{table}

\section{Applications and Results}
%non centrosymmetric Methylammonium ion could be opportinity to correctly determine space group in materials with centrosymmetric ions correct space group could be overseen

In hybrid halide perovskite semiconductors the presence of organic ions generally reduces the crystal symmetry compared to simple perovskites containing monoatomic ions. Therefore space groups listed in Table \ref{tab:classification} cannot be expected to occur there. However, some hybrid halide perovskites show the negative volume change across phase transitions, which can be viewed as a fingerprint of the dynamical tilting.

%%% corrections transferred

\subsection{Volume changes across phase transitions and negative thermal expansion}
Positive thermal expansion is common for many materials, a property which is maintained even across phase transitions and linked to extended thermal movement upon temperature increase or more precisely to lowering of the free energy upon lattice expansion. A negative thermal expansion is rather unusual and needs clarification.

The explanation of the negative thermal expansion comes naturally when dynamical tilting is considered. It should be mentioned that any (dynamic or static) freezing of a tilting increases the volume of the unit cell - as long as the octahedra are considered to be rigid, because within this approximation the volume is proportional to
\begin{eqnarray}  % \nonumber to remove numbering (before each equation)
\label{eqn:volu_angle}
\Phi=    \cos(a)^2  \cdot \cos (b)^2 \cdot \cos(c)^2   \; .
\end{eqnarray}
where $a,b$ and $c$ are tilting angles. $\Phi$ thus has a maximum at $a=b=c=0$, and therefore the cubic \emph{Pm}$\overline{3}$\emph{m} structure usually denoted as the  tilting system $a^0a^0a^0$ should have largest volume, which however is not always  observed contrary to geometric considerations. In the presence of dynamical tilting the cubic structure is rather attributed to the $ a^d a^d a^d  $ tilting system. Therefore the sharp structural phase transition at 54~$^\circ$C in CH$_3$NH$_3$PbI$_3$  from a tetragonal to the cubic phase  \cite{jacobsson2015} which is linked to a volume drop can be understood as an activation of additional dynamic tiltings leading to a volume decrease according to equation~\ref{eqn:volu_angle}. This explanation is further backed up by calculations which show large negative portions of the band structure around the $R$- and $M$-point of the Brillouin zone corresponding to octahedral tilting \cite{brivio2015} and observations revealing large thermal movement \cite{tyson2017}. \\

It should be mentioned similar observations are available in other structures e.g. in KNbO$_3$. There the volume change is positive across both phase transitions observed between 300 and 750~K \cite{sakakura2011}, with the cubic \emph{Pm}$\overline{3}$\emph{m} structure at the high temperature end of series of transitions.

\subsection{Agreement between lattice parameters and tilts system} %\subsection{\emph{Fmmm} and incompatibility of lattice parameters with tilting angles} %
\label{chap:fmmm}

The lattice parameters published e.g. by Brivio et al.~\cite{brivio2015} for methylammonium lead iodide are in agreement with the underlying tilt system. For the tetragonal \emph{I}4/\emph{mcm} phase the $c/a>1$ while for  the orthogonal phase the  calculated $c/a<1$  is consistent with their respective tilt system. In other materials such as Pr$_{0.5}$Sr$_{0.5}$MnO$_3$, which is reported in the \emph{Fmmm} and \emph{I}4/\emph{mcm} phase, static tilts cannot explain the pseudo-cubic $c/a$-ratio of 1.02 (tilt system $a^-b^0b^0$).
Static tilt would require rotation of more than 10$^\circ$, whereas the observed rotation angle is only 3.8$^\circ$. The consistent results can be obtained by considering dynamic tilts, which leads to modification of the lattice parameters without the change crystal coordinates resulting  in the \emph{Fmmm} space group.
% It undergoes a phase transition from the orthorhombic to the tetragonal phase \cite{caballero2014, damay1998} at $T_C$=135 -- 165~K. It thus represents a perovskite reported in a space group \emph{Fmmm} not compatible with the space groups resulting from static tilts in perovskites \cite{aleksandrov1976}. Considering dynamics tilts on the other hand, the tilting can be identified as $a^-b^dc^d$.
The \emph{Fmmm} structure is also reported for other Pr/Sr proportion \cite{knivzek2004} and for combined Pr-Sr-Ce doping \cite{heyraud2013}. The Pr-sites retain their statistically distributed fractional occupancies and no ionic ordering takes place.
It is tempting to explain the discrepancy between tilt system and cell parameters assuming distortion of octahedral sites. This however would undermine the success of the theory of static tilts based on rigid octahedra. Considering dynamical tilting it comes natural, that non-zero average tilt-amplitude reduces cell parameters perpendicular to the tilting axis, while as the instantaneous geometry of the octahedron remains rigid.

%According to Ref.~\cite{adams2016} it is a common to observe a multi-well PES as shown in Fig.~\ref{fig:states_single_well_double_well}~\textbf{a)} where the average position is shifted to $a\neq 0$. It can be considered as a superposition of dynamic tilts (Fig.~\ref{fig:states_single_well_double_well}~\textbf{a)}) with static tilts (see Fig.~\ref{fig:states_single_well_double_well}~\textbf{b)}). Because of the octahedral tilting of $a\neq 0$ (Fig.~\ref{fig:undist_octahedra}~\textbf{b)}), they follow Glazer's classification; however, when comparing the distortion of the pseudo-cubic axes calculated from octahedral tilting (i.e. the atomic X-positions) with the ones observed in experiment discrepancies appear. We can show this with the example of Pr$_{0.5}$Sr$_{0.5}$MnO$_3$ in the follows following.

\subsection{Instantaneous atomic positions and time averaged structure}

Egger et al. \cite{egger2018} discussed the possibility for `dynamical disorder' in halide perovskites (`dynamical tilting' in our terminology). The arguments for the presence of dynamic disorder were obtained analysing structural results obtained by complementary experimental methods. While X-diffraction yields an averaged structure with high symmetry, Raman shows a local structure with low symmetry. Similarly Beecher et al.~\cite{beecher2016} report ``anharmonic modes [...] with diffusive (order-disorder) dynamics persisting many tens to hundreds of Kelvin above the transition''. Still as has been stated, these modes are unobservable by Bragg diffraction. Indeed, these simultaneous observations can be reconciled through the concept of `dynamical tilting': We interpret these findings as snapshots of the dynamic tilting. It has been stated \cite{kassan1986} that the multi-well nature of the atomic potential energy surface cannot only lead to structural phase transitions to new phases which is the main subject of this paper but also to diffuse scattering: above the phase transition temperature the atoms retain some characteristics of the interactions below the phase transition (e.g. coupling of the octahedra), which leads to correlated movement of the octahedra.
\\
The  transition rate may vary, depending on the height of the energy barrier between adjacent potential energy minima,\footnote{This can be described in terms of Fermi's golden rule for the transition rate $\Gamma_{i\rightarrow f}= \frac {2\pi}{\hbar} \left | \langle f | V | i \rangle \right | ^2 \rho$, where $f$ and $i$ are final and the initial state respectively, $\rho$ the density of final states and $\langle f|V|i\rangle$ matrix element connecting the two states.}  High energy barrier for tilting can result in sluggish dynamics, at a time scale significantly larger than diffraction experiment, making visible the tilting of the octahedra (through the diffuse scattering) in crystals of otherwise higher symmetry. This high symmetry appears in the time averaged X-ray patters, or equivalently in the averaged structure in MD.
\\
% distingush between observed dynamic tilting and correlated tilting of the octahedra at T>T_C !
%However the short range ordering disappears in the distorted tilting while as in the dynamic tilting it can remain: when octahedra retain some interaction therefore move mostly in phase or out of phase, their correlation can be observed either in snapshots from MD, from ultra short time diffraction experiments, or even in normal diffraction experiments, if the dynamics of the tilting is slow.<

Similar observations can be made for other structures such as PbZrO$_3$ and Zr-rich PbZr$_{1-x}$Ti$_x$O$_3$, which are known to adopt a cubic $Pm{\rm \overline{3}}m$ structure above the Curie temperature of $T_C=$523~K \cite{zhang2015}.\footnote{At room temperature a centrosymmetric structure (space group $Pbam$) is observed, resulting from antiparallel displacements of the cations on the (110) planes and oxygen octahedral tilts of type $a^-a^-c^0$ \cite{glazer1993}.} At high temperature $T> 523$~K however the diffraction pattern contains a considerable amount of diffuse scattering, which can be attributed to distortion modes at the $M$-point in the Brillouin zone, i.e. correlated dynamical in phase tiltings along the crystal main axis in agreement with molecular dynamic simulations \cite{zhang2015}.

\subsection{Possible space groups for perovskites with dynamical tilting}

Often the structure refinement for perovskites is restricted to the space groups which are listed in the tables of Glazer \cite{glazer1972}, Aleksandrov \cite{aleksandrov1976}, or Howard and Stokes \cite{howard1998} and therefore can be explained by static tilts.

As mentioned already for hybrid halide perovskite structural refinement should not be restricted to these structures, due to the presence of an organic ions which alters crystal symmetry compared simple perovskites containing monoatomic ions. This has been widely accepted by research community working with hybrid halide perovskite and helps to avoid wrong crystal symmetry assignment and misinterpretation of phase diagrams.

It is described in the methods section, that dynamical tilting can lead to the space groups such as \emph{Fmmm}, \emph{P}4/\emph{mmm}, \emph{Cmmm} and \emph{Pmmm} which cannot be explained by static tilts. In section \ref{chap:fmmm} and in the discussion (\ref{sec:p4mmm} - \ref{sec:cryolite})
we review published perovskites structures with \emph{Fmmm}, \emph{P}4/\emph{mmm}, \emph{Cmmm} and \emph{Pmmm} symmetry and show that also in perovskite CaTiO$_3$ and cryolite $\rm Na_3AlF_6$ the known data point towards the presence of dynamical tiltings in these system.

%\section{Experimental evidence for dynamic tilting in perovskites}
\section{Discussion}\label{sec:exerpimental_evidence}

The dynamic tilting can explain the some observed structural phase transitions and important physical properties within perovskite class of structures.
In many compounds, the composition forbids an atomic arrangement corresponding precisely to the perovskite ABX$_3$ described above. This is often due to the chemical composition involving further chemical species or the ordering of the ions, frequently occurring  in a rock-salt like arrangement of the $A$ and $B$ cations or of the involved octahedra.
Many of the resulting compounds are captured by the structure formula A$_2$B'B''X$_6$ and are called double-perovskites or layered perovskites. This class retains the stability and often also the rigidity of the octahedra \cite{hossain2018}, while as more chemical compositions are feasible than in the simple ABX$_3$ composition. The layered perovskites are of great importance due to their strong and unusual magnetic interactions \cite{bristowe2015}, superconductivity \cite{bednorz1986} and technical applications \cite{fan2015, fan2015, granados2020}. We will therefore include some of these double perovskites in the discussion.

\subsection{P4/mmm} \label{sec:p4mmm} % \label{sec:cryolite}%

Many structures are reported in the \emph{P}4/\emph{mmm} crystal structure,
BaTiO$_3$ % 3.9998(8) 3.9998(28) 4.0180
\cite{buttner1992} --- the name of which is used for the whole crystal class of perovskites with \emph{P}4/\emph{mmm} symmetry --- KCuF$_3$ and KCrF$_3$ \cite{edwards1959}, %  	 a,b,c=4.13 4.13 3.92 , room temperature, Medvedeva prb 65, 172413
CeAlO$_3$ \cite{tanaka1993}, % 3.7669(9) 3.7669(9) 3.7967, Vasylechko rports I4/mcm as groundstate, but the phase diagramm I4/mcm/imma/r3c/pm-3m is stange
TlCuF$_3$ \cite{rudorff1963},
SrFeO$_3$ \cite{diodati2012}, % 3.854 3.854 3.869
CsAuCl$_3$ \cite{matsushita2007}  % 4.8196(8) 4.8196(8) 4.8293
and CeGaO$_3$ \cite{shishido1997}. % 3.873(1) 3.873(1) 3.880

For BaTiO$_3$ the phase transitions from rhombohedral to orthorhombic, tetragonal and cubic is known \cite{hayward2002}. It indicates, that the \emph{P}4/\emph{mmm} is stabilized by entropy and that the PES shows a negative curvature.
We would like to stress here, that dynamic instabilities do not necessarily indicate a structural instability at $T=0$~K. In general the correct ground state energy is the sum of the minimum of the potential energy $E_0$ and the zero-point energy $E_{\rm ZP}$. The calculations suggest that $E_{\rm ZP}$ lies particularly low for in structures with a multi-well potential-energy surface. In some structures --- especially those exhibiting a multi-well potential-energy --- $E_0+E_{\rm ZP}$ can be lower at a saddle point of the PES than at the structure relaxed at a minimum of the potential energy. This could potentially be overseen by atomistic simulations, where often only $E_0$ is accounted for. Therefore, for the remaining structures reported (CeAlO$_3$, TlCuF$_3$, SrFeO$_3$, CsAuCl$_3$, CeGaO$_3$) the stabilization through the zero-point energy is plausible, as they remain stable in the \emph{P}4/\emph{mmm} phase down to low temperatures.

\subsection{Cmmm }

The \emph{Cmmm} space group has been reported for (Li,La)TiO$_3$-perovskite-systems which in turn give its name to the crystal class \cite{sanz2004}. \\
NaIO$_3$ is also known in the \emph{Cmmm} space group \cite{zachariasen1928}, as well as  Nd$_{0.7}$TiO$_3$ \cite{sefat2005}.

\subsection{Pmmm}

Another crystal structure type is \emph{Pmmm} PbTiO$_3$-perovskite \cite{cole1937}.  At ambient pressure it undergoes a phase transition to the cubic phase between 600 and 800~K \cite{zhu2011}. The same space group is also found for NaNbO$_3$ \cite{solov1961}, Mg$_{0.5}$W$_{0.5}$O$_3$ \cite{zaslavskii1963} and GdCoO$_3$ \cite{ruggiero1954}.
\\
The NaNbO$_3$ attracts particularly our attention: It has been refined recently \cite{peel2012} for the a high temperature phase at 500$^\circ$C. Best refinements were achieved with space group \emph{Pmmm} ($\chi^2=1.80$) and \emph{Pnma} ($\chi^2=1.85$).

% Checked, PbTiO3 is ferro-electric, but the structure shows now cation displacement (

\subsection{Perovskite CaTiO$_3$}

In CaTiO$_3$ a cascade of phase transitions from \emph{Pbnm} , to \emph{Cmcm} (1380~K), \emph{I}4/\emph{mcm} (1500~K) and finally to the cubic \emph{Pm}$\overline{3}$\emph{m} phase (1580~K) is observed. For the first transition  (\emph{Pbnm} to \emph{Cmcm} at 1380~K) small anomalies in the temperature dependence of the cell and structural parameters are observed \cite{kennedy1999}.\\

However, other authors mention the large atomic-displacement parameter of the cubic phase and state: ``the high-temperature phase transition to cubic perovskite is triggered by the sudden increase of the mobility of the oxygen sublattice or at least of parts of it.'' \cite{vogt1993}. Quenching allows to access the dynamic disorder of the cubic phase \cite{britvin2022}. Furthermore, within the approximation of static tilts the fading of tilting angles at the \emph{I}4/\emph{mcm} -\emph{Pm}$\overline{3}$\emph{m} phase boundary at 1500~K (\cite{kennedy1999}; Fig. 4) should lead to a significant increase of the volume in a temperature range of 20-50~K, which is not observed. Therefore the cubic phase should be assigned to a dynamic tilting of oxygen octahedra $a^da^da^d$ rather to an ideal perovskite structure $a^0a^0a^0$.

\subsubsection{Cryolite $\rm Na_3AlF_6$} \label{sec:cryolite}%

\begin{figure}
 \includegraphics[width=0.98\textwidth]{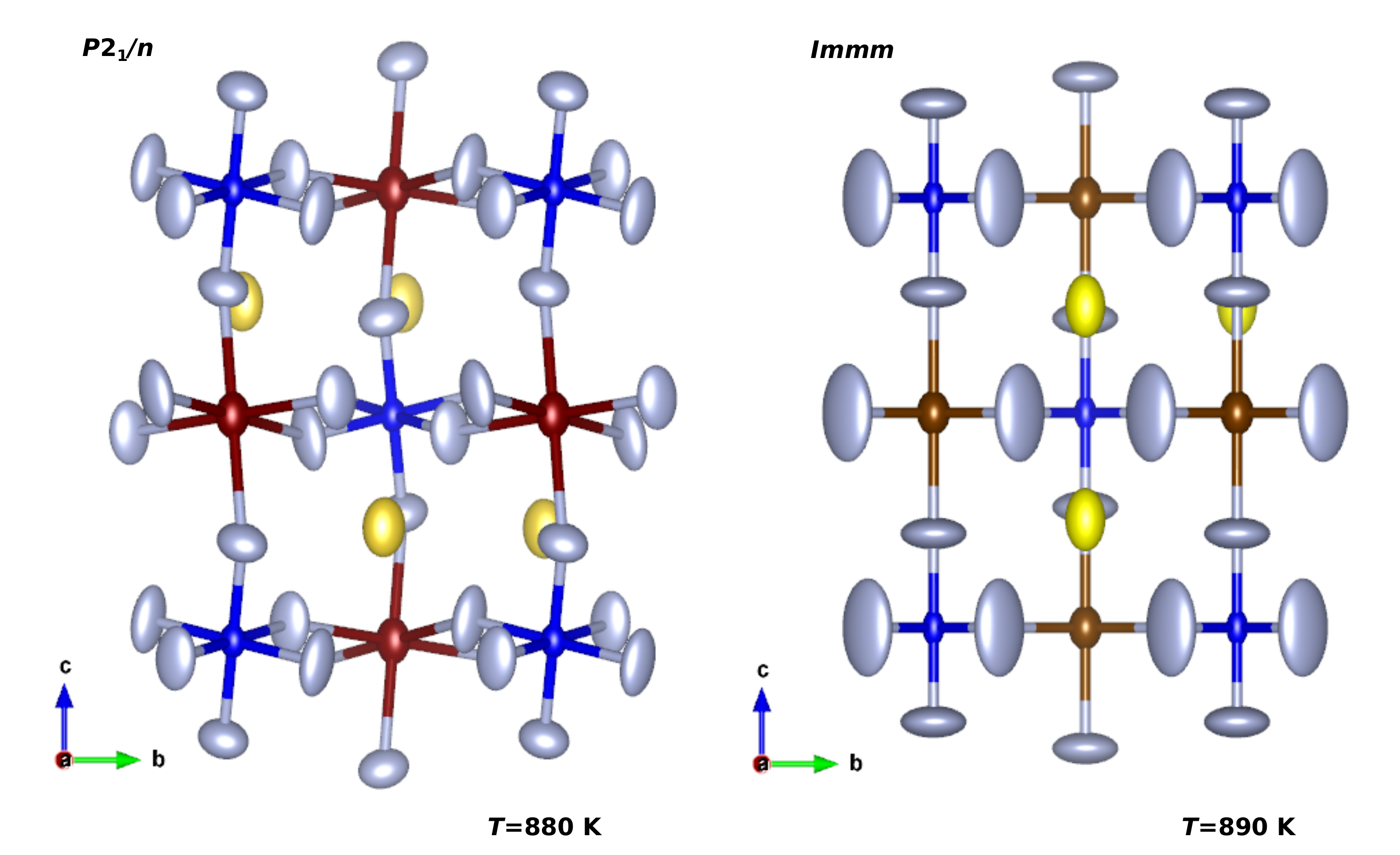}\\
 \caption{Thermal vibrational ellipsoids in cryolite at $T=880$~K and $T=890$~K, i.e. below and above the phase transition as reported in Ref.~\cite{yang1993}. Color code: Al blue, Na1 and Na2 blue and yellow, F grey.
% $\rm Na_2 Na Al F_6$
 At $T=890$~K in the high temperature phase neither the atomic positions nor the vibrational octahedra suggest static tilting.}\label{fig:yang_vib_octahedra}
% $a^0b^+c^+$ which according to the theory of static tilting could explain the \emph{Immm} symmetry.
 % (or golden at $T=890$~K)
\end{figure}

The phase transition between the \textit{P2}$_1/$\textit{n} and the \textit{Immm} space group \cite{anthony2005} is well documented in cryolite. The critical temperature is $T_c=885$~K (612~$^\circ$C). Due to excellent X-Ray data the atomic positions and the main axes of the vibrational ellipsoids are known \cite{yang1993} below and above $T_C$. \\
At low temperature, the systems shows static tilts. The \textit{P2}$_1/$\textit{n} space-group is generated by rigid octahedra and a tilt system with one in-phase tilting and two out-of-phase tiltings --- see also Fig.~\ref{fig:yang_vib_octahedra}.
Due to the rock-salt ordering of the octahedra the classification for simple perovskites \cite{lufaso2001, woodward1997a, woodward1997b} --- where $a^+b^-c^-$ corresponds to \textit{P2}$_1/$\textit{m}  --- cannot be applied.
%(tilt systems 8 or 9 in the classification of Woodward)
\\
The experimental high temperature orthorhombic \textit{Immm} structure shows displacements of fluorine from the ideal cubic perovskite positions. By further inspection they result from diffrent bonding length in AlO$_6$ and NaO$_6$ octahedra. The pseudo-cubic lattice parameters on the other hand suggest a tilting by at least 2.5$^\circ$
%, i.e. the lattice is orthorhombic, which is incompatible with the concept of rigid octahedra.
Finally the main axes of vibrational ellipsoids can increase by more than 200\% over the phase transition. All this points toward a system with at least one \emph{dynamic} tilting.
% s e.g. $a^+ b^d c^+$ also corresponding to the \textit{Immm} space group.
% consider also cell parameters (Tilt systems 1, 2 or 15 indicate c/a>1 ? while as we have c/a<1

% Checked, PbTiO3 is ferro-electric, but the structure shows no cation displacement (

\subsection{Impact of dynamical tiling on the physical properties}
% Consequences on the physical properties of dynamic tiling for perovskites

% Beecher but can explain key material properties such as the structural phase sequence, ultralow thermal transport, and large minority charge carrier lifetimes despite moderate carrier mobility.

The multi-well energy landscape and the amplification of thermal movement upon temperature increase lead to
\begin{enumerate}%[leftmargin=*,labelsep=4.9mm]
 \item unusually large thermal anisotropic displacement factors for B and X sites (see e.g. Ref.~\cite{tyson2017} for CH$_3$NH$_3$PbI$_3$)
 \item negative (directional) temperature expansion (see e.g. Ref.~\cite{jacobsson2015} for the negative directional temperature expansion in CH$_3$NH$_3$PbI$_3$)
 \item and volume decrease at the phase boundary upon temperature increase (see e.g. Ref.~\cite{sakakura2011} discussing the volume collapse in Na$_{0.5}$K$_{0.5}$NbO$_3$ at the phase boundaries at 446 and 666~K).
\end{enumerate}

% from energy perspective: dynamic tilting has many low lying energy excitations (even degenerate)
% After: Filip et al.~\cite{filip2018}  properties $\rm MAPbI_3$ calculation and experiment ; considering the electron-phonon interaction:
Wright et al.~\cite{wright2016} investigate the electron-phonon in hybrid lead halide perovskites in order to access the coupling transport properties, charge carrier recombination and finally the charge carrier mobility. It is surprising that the structural instabilities are overseen in these calculations: The harmonic approximation result in imaginary frequencies in these materials (see also Ref.~\cite{yang2017acoustic}) which is a signature of the negative curvature of the PES. However, the frequencies cannot be interpreted physically. Adams and Passerone~\cite{adams2016} showed, that dynamic instabilities can lead to high DOS in the phonon dispersion relation around $\omega=0$. Furthermore the electron-phonon coupling coefficient depends in the atomic mean square deviation (MSD) \cite{antonius2015}, which diverges for unstable modes in harmonic theory, and which is overestimated for stable modes in harmonic theory \cite{adams2020}. It thus remains to be seen how the reevaluation of the electron-phonon matrix element will change our picture of the electron-phonon in hybrid lead halide perovskites. \\
%% distortion from orbital ordering in KCuF$_3$ \cite{edwards1959}
%They could couple to the electronic states at the Fermi-energy and thus modify the electron-electron interaction in the solid. They therefore could significantly contribute to the interesting electronic phenomena e.g. in KCuF$_3$ \cite{edwards1959}.

However, the importance of anharmonic vibrational excitations in stabilisation of the different phases of halide perovskites are known. They are accessed through Monte Carlo simulations (see e.g.~\cite{bechtel2019} for CsPbBr$_3$) or Landau theory, which are computationally costly or semi-empirical, respectively (e.g. for CsPbI$_3$ \cite{marronnier2017}). The evaluation of the correct vibrational spectrum e.g. in DAMA \cite{adams2006, adams2020} however, gives access to the free energy and thus to most physical and thermodynamic properties of the material. In some structures the formation of dynamic instabilities seems to be fostered by pressure e.g. in CsAuCl$_3$ \cite{matsushita2007}. In these structure the evaluation of the free energy could clarify the role of pressure in the structure stabilisation.

As mentioned in the introduction, the multi-well character of the PES results in a high sensitivity of the ionic positions on perturbations. This is also reflected in the electron-phonon coupling constant
\[ \alpha _F\propto \sqrt{\frac{m}{2\hbar \omega} }\]
were $\hbar$ is Planck's reduced constant, $m$ is the effective mass of the charge carrier and and $\omega$ is the LO phonon frequency \cite{feynman1955}. For degenerate eigenstates the transition frequency vanishes and the coupling therefore diverges. Therefore the concept of polarons has to be extended.

\section{Conclusions}
%\subsection{Group/sub group relations}

\subsection*{Phase stabilisation}
The dynamic tilts can set in upon temperature increase. Their mechanism can be explained based on our results in Ref.~\cite{adams2016}. The double well potential energy surface is the necessary condition for the onset of dynamic tilt (Fig.~\ref{fig:states_single_well_double_well}). The energies of the vibrational modes in the double well potential are more dense (a) than in a single well (b) (compare also the scale of the energy axis). Most importantly, the ground state of the double well is almost degenerate, leading to a particularly high density of states at the lowest energy \footnote{This high density is calculated in Ref.~\cite{adams2016} for the vibrational DOS of cryolite.}. Typically the high temperature phase lies energetically higher than the low temperature phase, e.g. $\Delta E= 90$~meV per $\rm CH_3NH_3PbCl_3$ unit has been reported for the cubic phase compared to the its orthorhombic phase \cite{brivio2015}. This energy difference is outweighed by the relevant thermodynamic potential at temperature, which is the free energy
\begin{equation}\label{eqn:free_energy}
   A= -k_B\ln( \sum_i e^{-\frac{\epsilon_i}{k_B T}})
\end{equation}
with $k_B$ the Boltzmann-constant, $T$ the temperature and $\epsilon_i$ the energy of the vibrational excitation. $A$ decreases whenever the vibrational energies $\epsilon_i$ decrease, e.g. through lattice expansion, or here when a phase transition leads to a high density of states at low excitation energies e.g. through degenerate low lying modes in the cubic phase of CsPbI$_3$  \cite{marronnier2017}.

The activation energy i.e. the energy required to surmount the energy barrier between different energy minima - see Fig.~\ref{fig:states_single_well_double_well} a), is, in general comparable with the energy of thermal vibrations at the transition temperature. It is particularly low for structures with small tilting, which according to Glazer \cite{glazer1972} appears mostly in structures where the size of the A cation matches the one of the cavity. This leads to a small tilting and thus a low energy barrier with relatively low transition temperatures (e.g. 7.3~meV in CsPbI$_3$ in the cubic phase which transforms  to the tetragonal $\beta$-phase at 260~$^\circ$C \cite{marronnier2017}). If the A cation is small (e.g. MgSiO$_3$) the tilting is larger, and the phase transition can potentially occur at higher temperatures. This kind of transition can be expected to be very common. In some systems, however, the critical temperature lies above the melting point and therefore such transition is not observed.

\subsection*{Group-subgroup relations}
It should be emphasised that in general the appearance or disappearance of a dynamic tilting mode does not lead to the simple group/sub-group relation between resulting structures. This can be shown in the example of the
\[ Pnma \; (a^+b^-b^-) \; \rightarrow\;  Cmcm \; (a^db^+c^-)  \; \] % SGN 62 to 63
transition, where one $b^-$ transforms to the $a^d$ dynamic tilt. Considering group-subgroup relations at least one intermediate subgroup is involved (either \emph{Pbcm} or \emph{Pmmm}) \footnote{Data from Bilbao crystallographic server \cite{aroyo2006}.}.
In other structures the onset of dynamic tilt does not lead to space group change, e.g. in the case of
\[ Imma \; (a^-a^-b^0) \; \rightarrow\;  Imma \; (a^-a^-b^d)  \; \] %
The shortening of the crystal axis related to the onset of the dynamic tilt does not break any symmetry. As shown in these two examples dynamic tilts do not correspond to irreducible representations of the space group and the two end members are related \emph{in a simple or more complex way.}

%\subsection{General scheme what observed structures are likely candidates}
\subsection*{Crystallographic consequences of dynamical tiling}
As shown in the few examples the particular structural relationships in the structures at phase transition can be indicative transformations driven by dynamic tilts. Dynamic tilt should be considered if:
\begin{enumerate}%[leftmargin=*,labelsep=4.9mm]
  \item the observed instantaneous symmetry does not correspond to the average symmetry.
   \item the experimentally observed space group of a perovskite-type structure is not listed in the tables established by Glazer \cite{glazer1972}, Woodward \cite{woodward1997a} or Aleksandrov \cite{aleksandrov1976}.
  \item the proportions of the lattice parameters are at odds with the ones indicated from the theory of static tilts. E.g. in the $I4/mcm$ symmetry, which appears in the $(a^da^db^-)$ as well as in the $(a^0a^0b^-)$ tilting system, the first one corresponds strictly to a lattice ratio of $c/a>1$, while as the second one has no restrictions for this ratio.
 \item the octahedra appear distorted. The octahedra represent a stable configuration of the BX$_6$ chemical configuration. (Large) distortions are unlikely, except in the case of Jahn-Teller distortions. The success of the theory of static tilts in numerous other systems \cite{aleksandrov1976, aleksandrov1994, glazer1972, woodward1997a, woodward1997b} underlines the correctness of these considerations.
\end{enumerate}

\subsection*{Dynamical tilting and electron-phonon interaction}
The interaction between phonons and charge carriers is an important for understand the physical mechanisms controlling the mobility of the charge carriers. High mobility is desirable for applications which are based on the efficient separation of electrons and holes, e.g. in photovoltaics. It is significantly influenced by the interaction between photons and the charge carriers. Steele et al.~\cite{steele2019} confirm a strong electron-phonon coupling in $\rm CH_3NH_3PbCl_3$ and relate it to `strong anharmonicity and dynamic disorder'. Similar results are found for $\rm CsPbBr_3$ and $\rm CH_3NH_3PbBr_3$ \cite{sendner2016}. The the interdependence between dynamic structural distortions, photo-carriers and photons in lead halide perovskites is also documented in Ref.~\cite{batignani2018}. The effect of this coupling on the mobility needs further investigation as at large coupling its temperature dependence can exhibit multiple extrema \cite{prodanovic2019}.
Such a coupling can have further unexpected effects on the photoluminescence, such as the up-conversion i.e. an increase of the luminescence frequency through energy transfer from phonons \cite{granados2020}.
%  In metal halide perovskites they are described as collective modes of charge carriers and rotations of the organic cation sublattice \cite{ghosh2020}.

Based on available observations, we can conclude, that dynamical tilting can affect following properties in perovskites, especially in lead halide perovskites:
\begin{itemize}
  \item the electron-phonon coupling. The electron-phonon coupling coefficient depends in the atomic MSD, which is significantly increased in structures showing dynamical tilting compared to structures without structural instabilities.
  \item the charge carrier mobility. Amongst others it is limited by the interaction of chage-carriers with crystal vibrations \cite{herz2017}. This interaction can be described in terms of polarons, which consists of the polarisation of the ionic lattice by a mobile electron \cite{frohlich1954, feynman1955}. The resulting field finally creates e.g. the Frohlich interaction. As mentioned above, in a structure with dynamic tiltings the ionic eigenstates of a double-well of the PES are degenerate, which results in high sensitivity of the ionic positions on perturbations, and presumably large electron-phonon coupling constant and polarizability.
  \item the spectral width of light-emitting semiconductor devices \cite{iaru2017} by electron-phonon interactions. Wright et al.~\cite{wright2016} underline, that the ``Frohlich coupling to LO phonons is the predominant charge-carrier scattering mechanism in hybrid lead halide perovskites'', leading to emission linewidth broadening --- see also Ref.~\cite{herz2017}.
\end{itemize}

\hevo{All these properties seem to be tightly related to the observed instability on the surface of the potential energy and the resulting dynamical tilting. Therefore the inherent structural sensibility might be the premise of the high efficiency of perovskite solar cells, which indicates that the challenge in the application of these materials is the stabilisation of the structure against phase transitions. Temperature stabilisation of $\rm CSPbI_3$ \cite{kirschner2019} points to stabilisation through the free energy (see also equation \ref{eqn:free_energy}) and thus confirms our model.}{A3}

Many perovskite structures show interesting properties such as photoelectricity, but also magnetism, ferroelectricity or superconductivity. The dynamically tiltings result in almost degenerate phonon vibrational modes \cite{adams2016} and worth further investigations. These modes could couple to the electronic states and thus modify the electron-electron interaction in the solid in an unexpected many. For photoelectric materials Marronier et al.~\cite{marronnier2017} explicitly state that `the perovskite oscillations through the corresponding energy barrier could explain the underlying ferroelectricity and the dynamical Rashba effect predicted in halide perovskites for photovoltaics''.	
A number of physical properties of perovskite structures are yet to be explained. The space groups for dynamic tilting of cubic perovskties reported in this work will facilitate consideration of dynamical tilting in connection with physical properties of perovskites.

\section*{Acknowledgments}
This research was supported by a grant from the Swiss National Supercomputing Centre (CSCS) under project ID s792.
 The authors thank Rene Schliemann for help with geometry determination of the structure and Georgia Cametti for visualization. We are particularly grateful to Prof. Thomas Armbruster for stimulating discussions and careful reading of the manuscript.

\bibliography{bibn}

\begin{thebibliography}{82}%
\makeatletter
\providecommand \@ifxundefined [1]{%
 \@ifx{#1\undefined}
}%
\providecommand \@ifnum [1]{%
 \ifnum #1\expandafter \@firstoftwo
 \else \expandafter \@secondoftwo
 \fi
}%
\providecommand \@ifx [1]{%
 \ifx #1\expandafter \@firstoftwo
 \else \expandafter \@secondoftwo
 \fi
}%
\providecommand \natexlab [1]{#1}%
\providecommand \enquote  [1]{``#1''}%
\providecommand \bibnamefont  [1]{#1}%
\providecommand \bibfnamefont [1]{#1}%
\providecommand \citenamefont [1]{#1}%
\providecommand \href@noop [0]{\@secondoftwo}%
\providecommand \href [0]{\begingroup \@sanitize@url \@href}%
\providecommand \@href[1]{\@@startlink{#1}\@@href}%
\providecommand \@@href[1]{\endgroup#1\@@endlink}%
\providecommand \@sanitize@url [0]{\catcode `\\12\catcode `\$12\catcode
  `\&12\catcode `\#12\catcode `\^12\catcode `\_12\catcode `\%12\relax}%
\providecommand \@@startlink[1]{}%
\providecommand \@@endlink[0]{}%
\providecommand \url  [0]{\begingroup\@sanitize@url \@url }%
\providecommand \@url [1]{\endgroup\@href {#1}{\urlprefix }}%
\providecommand \urlprefix  [0]{URL }%
\providecommand \Eprint [0]{\href }%
\providecommand \doibase [0]{http://dx.doi.org/}%
\providecommand \selectlanguage [0]{\@gobble}%
\providecommand \bibinfo  [0]{\@secondoftwo}%
\providecommand \bibfield  [0]{\@secondoftwo}%
\providecommand \translation [1]{[#1]}%
\providecommand \BibitemOpen [0]{}%
\providecommand \bibitemStop [0]{}%
\providecommand \bibitemNoStop [0]{.\EOS\space}%
\providecommand \EOS [0]{\spacefactor3000\relax}%
\providecommand \BibitemShut  [1]{\csname bibitem#1\endcsname}%
\let\auto@bib@innerbib\@empty
%</preamble>
\bibitem [{\citenamefont {Bednorz}\ and\ \citenamefont
  {M{\"u}ller}(1986)}]{bednorz1986}%
  \BibitemOpen
  \bibfield  {author} {\bibinfo {author} {\bibfnamefont {J.}~\bibnamefont
  {Bednorz}}\ and\ \bibinfo {author} {\bibfnamefont {K.}~\bibnamefont
  {M{\"u}ller}},\ }\href@noop {} {\bibfield  {journal} {\bibinfo  {journal}
  {Zeitschrift f{\"u}r Physik B Cond. Matt.}\ }\textbf {\bibinfo {volume}
  {64}},\ \bibinfo {pages} {189} (\bibinfo {year} {1986})}\BibitemShut
  {NoStop}%
\bibitem [{\citenamefont {Fan}\ \emph {et~al.}(2015)\citenamefont {Fan},
  \citenamefont {Xiao}, \citenamefont {Sun}, \citenamefont {Chen},
  \citenamefont {Hu}, \citenamefont {Ouyang}, \citenamefont {Ong},
  \citenamefont {Zeng},\ and\ \citenamefont {Wang}}]{fan2015}%
  \BibitemOpen
  \bibfield  {author} {\bibinfo {author} {\bibfnamefont {Z.}~\bibnamefont
  {Fan}}, \bibinfo {author} {\bibfnamefont {J.}~\bibnamefont {Xiao}}, \bibinfo
  {author} {\bibfnamefont {K.}~\bibnamefont {Sun}}, \bibinfo {author}
  {\bibfnamefont {L.}~\bibnamefont {Chen}}, \bibinfo {author} {\bibfnamefont
  {Y.}~\bibnamefont {Hu}}, \bibinfo {author} {\bibfnamefont {J.}~\bibnamefont
  {Ouyang}}, \bibinfo {author} {\bibfnamefont {K.~P.}\ \bibnamefont {Ong}},
  \bibinfo {author} {\bibfnamefont {K.}~\bibnamefont {Zeng}}, \ and\ \bibinfo
  {author} {\bibfnamefont {J.}~\bibnamefont {Wang}},\ }\href@noop {} {\bibfield
   {journal} {\bibinfo  {journal} {J. Phys. Chem. Lett.}\ }\textbf {\bibinfo
  {volume} {6}},\ \bibinfo {pages} {1155} (\bibinfo {year} {2015})}\BibitemShut
  {NoStop}%
\bibitem [{\citenamefont {Granados~del {\'A}guila}\ \emph
  {et~al.}(2020)\citenamefont {Granados~del {\'A}guila}, \citenamefont {Do},
  \citenamefont {Xing}, \citenamefont {Jee}, \citenamefont {Khurgin},\ and\
  \citenamefont {Xiong}}]{granados2020}%
  \BibitemOpen
  \bibfield  {author} {\bibinfo {author} {\bibfnamefont {A.}~\bibnamefont
  {Granados~del {\'A}guila}}, \bibinfo {author} {\bibfnamefont {T.~T.~H.}\
  \bibnamefont {Do}}, \bibinfo {author} {\bibfnamefont {J.}~\bibnamefont
  {Xing}}, \bibinfo {author} {\bibfnamefont {W.~J.}\ \bibnamefont {Jee}},
  \bibinfo {author} {\bibfnamefont {J.~B.}\ \bibnamefont {Khurgin}}, \ and\
  \bibinfo {author} {\bibfnamefont {Q.}~\bibnamefont {Xiong}},\ }\href@noop {}
  {\bibfield  {journal} {\bibinfo  {journal} {Nano Research}\ }\textbf
  {\bibinfo {volume} {13}},\ \bibinfo {pages} {1962} (\bibinfo {year}
  {2020})}\BibitemShut {NoStop}%
\bibitem [{\citenamefont {Li}\ \emph {et~al.}(2018)\citenamefont {Li},
  \citenamefont {Zhang}, \citenamefont {Song}, \citenamefont {Xie},
  \citenamefont {Meng},\ and\ \citenamefont {Xu}}]{li2018}%
  \BibitemOpen
  \bibfield  {author} {\bibinfo {author} {\bibfnamefont {S.}~\bibnamefont
  {Li}}, \bibinfo {author} {\bibfnamefont {C.}~\bibnamefont {Zhang}}, \bibinfo
  {author} {\bibfnamefont {J.-J.}\ \bibnamefont {Song}}, \bibinfo {author}
  {\bibfnamefont {X.}~\bibnamefont {Xie}}, \bibinfo {author} {\bibfnamefont
  {J.-Q.}\ \bibnamefont {Meng}}, \ and\ \bibinfo {author} {\bibfnamefont
  {S.}~\bibnamefont {Xu}},\ }\href@noop {} {\bibfield  {journal} {\bibinfo
  {journal} {Crystals}\ }\textbf {\bibinfo {volume} {8}},\ \bibinfo {pages}
  {220} (\bibinfo {year} {2018})}\BibitemShut {NoStop}%
\bibitem [{\citenamefont {Bechtel}\ \emph {et~al.}(2019)\citenamefont
  {Bechtel}, \citenamefont {Thomas},\ and\ \citenamefont {Van~der
  Ven}}]{bechtel2019}%
  \BibitemOpen
  \bibfield  {author} {\bibinfo {author} {\bibfnamefont {J.~S.}\ \bibnamefont
  {Bechtel}}, \bibinfo {author} {\bibfnamefont {J.~C.}\ \bibnamefont {Thomas}},
  \ and\ \bibinfo {author} {\bibfnamefont {A.}~\bibnamefont {Van~der Ven}},\
  }\href@noop {} {\bibfield  {journal} {\bibinfo  {journal} {Phys. Rev.
  Materials}\ }\textbf {\bibinfo {volume} {3}},\ \bibinfo {pages} {113605}
  (\bibinfo {year} {2019})}\BibitemShut {NoStop}%
\bibitem [{\citenamefont {Zhao}\ \emph {et~al.}(2016)\citenamefont {Zhao},
  \citenamefont {Shi}, \citenamefont {Xi}, \citenamefont {Wang},\ and\
  \citenamefont {Shuai}}]{zhao2016}%
  \BibitemOpen
  \bibfield  {author} {\bibinfo {author} {\bibfnamefont {T.}~\bibnamefont
  {Zhao}}, \bibinfo {author} {\bibfnamefont {W.}~\bibnamefont {Shi}}, \bibinfo
  {author} {\bibfnamefont {J.}~\bibnamefont {Xi}}, \bibinfo {author}
  {\bibfnamefont {D.}~\bibnamefont {Wang}}, \ and\ \bibinfo {author}
  {\bibfnamefont {Z.}~\bibnamefont {Shuai}},\ }\href@noop {} {\bibfield
  {journal} {\bibinfo  {journal} {Scientific reports}\ }\textbf {\bibinfo
  {volume} {6}},\ \bibinfo {pages} {1} (\bibinfo {year} {2016})}\BibitemShut
  {NoStop}%
\bibitem [{\citenamefont {Wright}\ \emph {et~al.}(2016)\citenamefont {Wright},
  \citenamefont {Verdi}, \citenamefont {Milot}, \citenamefont {Eperon},
  \citenamefont {P{\'e}rez-Osorio}, \citenamefont {Snaith}, \citenamefont
  {Giustino}, \citenamefont {Johnston},\ and\ \citenamefont
  {Herz}}]{wright2016}%
  \BibitemOpen
  \bibfield  {author} {\bibinfo {author} {\bibfnamefont {A.~D.}\ \bibnamefont
  {Wright}}, \bibinfo {author} {\bibfnamefont {C.}~\bibnamefont {Verdi}},
  \bibinfo {author} {\bibfnamefont {R.~L.}\ \bibnamefont {Milot}}, \bibinfo
  {author} {\bibfnamefont {G.~E.}\ \bibnamefont {Eperon}}, \bibinfo {author}
  {\bibfnamefont {M.~A.}\ \bibnamefont {P{\'e}rez-Osorio}}, \bibinfo {author}
  {\bibfnamefont {H.~J.}\ \bibnamefont {Snaith}}, \bibinfo {author}
  {\bibfnamefont {F.}~\bibnamefont {Giustino}}, \bibinfo {author}
  {\bibfnamefont {M.~B.}\ \bibnamefont {Johnston}}, \ and\ \bibinfo {author}
  {\bibfnamefont {L.~M.}\ \bibnamefont {Herz}},\ }\href@noop {} {\bibfield
  {journal} {\bibinfo  {journal} {Nature Communications}\ }\textbf {\bibinfo
  {volume} {7}},\ \bibinfo {pages} {1} (\bibinfo {year} {2016})}\BibitemShut
  {NoStop}%
\bibitem [{\citenamefont {Neukirch}\ \emph {et~al.}(2016)\citenamefont
  {Neukirch}, \citenamefont {Nie}, \citenamefont {Blancon}, \citenamefont
  {Appavoo}, \citenamefont {Tsai}, \citenamefont {Sfeir}, \citenamefont
  {Katan}, \citenamefont {Pedesseau}, \citenamefont {Even}, \citenamefont
  {Crochet} \emph {et~al.}}]{neukirch2016}%
  \BibitemOpen
  \bibfield  {author} {\bibinfo {author} {\bibfnamefont {A.~J.}\ \bibnamefont
  {Neukirch}}, \bibinfo {author} {\bibfnamefont {W.}~\bibnamefont {Nie}},
  \bibinfo {author} {\bibfnamefont {J.-C.}\ \bibnamefont {Blancon}}, \bibinfo
  {author} {\bibfnamefont {K.}~\bibnamefont {Appavoo}}, \bibinfo {author}
  {\bibfnamefont {H.}~\bibnamefont {Tsai}}, \bibinfo {author} {\bibfnamefont
  {M.~Y.}\ \bibnamefont {Sfeir}}, \bibinfo {author} {\bibfnamefont
  {C.}~\bibnamefont {Katan}}, \bibinfo {author} {\bibfnamefont
  {L.}~\bibnamefont {Pedesseau}}, \bibinfo {author} {\bibfnamefont
  {J.}~\bibnamefont {Even}}, \bibinfo {author} {\bibfnamefont {J.~J.}\
  \bibnamefont {Crochet}},  \emph {et~al.},\ }\href@noop {} {\bibfield
  {journal} {\bibinfo  {journal} {Nano lett.}\ }\textbf {\bibinfo {volume}
  {16}},\ \bibinfo {pages} {3809} (\bibinfo {year} {2016})}\BibitemShut
  {NoStop}%
\bibitem [{\citenamefont {Karakus}\ \emph {et~al.}(2015)\citenamefont
  {Karakus}, \citenamefont {Jensen}, \citenamefont {DAngelo}, \citenamefont
  {Turchinovich}, \citenamefont {Bonn},\ and\ \citenamefont
  {Canovas}}]{karakus2015}%
  \BibitemOpen
  \bibfield  {author} {\bibinfo {author} {\bibfnamefont {M.}~\bibnamefont
  {Karakus}}, \bibinfo {author} {\bibfnamefont {S.~A.}\ \bibnamefont {Jensen}},
  \bibinfo {author} {\bibfnamefont {F.}~\bibnamefont {DAngelo}}, \bibinfo
  {author} {\bibfnamefont {D.}~\bibnamefont {Turchinovich}}, \bibinfo {author}
  {\bibfnamefont {M.}~\bibnamefont {Bonn}}, \ and\ \bibinfo {author}
  {\bibfnamefont {E.}~\bibnamefont {Canovas}},\ }\href@noop {} {\bibfield
  {journal} {\bibinfo  {journal} {J. Phys. Chem. Lett.}\ }\textbf {\bibinfo
  {volume} {6}},\ \bibinfo {pages} {4991} (\bibinfo {year} {2015})}\BibitemShut
  {NoStop}%
\bibitem [{\citenamefont {Samara}\ \emph {et~al.}(1990)\citenamefont {Samara},
  \citenamefont {Hammetter},\ and\ \citenamefont {Venturini}}]{samara1990}%
  \BibitemOpen
  \bibfield  {author} {\bibinfo {author} {\bibfnamefont {G.}~\bibnamefont
  {Samara}}, \bibinfo {author} {\bibfnamefont {W.}~\bibnamefont {Hammetter}}, \
  and\ \bibinfo {author} {\bibfnamefont {E.}~\bibnamefont {Venturini}},\
  }\href@noop {} {\bibfield  {journal} {\bibinfo  {journal} {Phys. Rev. B}\
  }\textbf {\bibinfo {volume} {41}},\ \bibinfo {pages} {8974} (\bibinfo {year}
  {1990})}\BibitemShut {NoStop}%
\bibitem [{\citenamefont {Beecher}\ \emph {et~al.}(2016)\citenamefont
  {Beecher}, \citenamefont {Semonin}, \citenamefont {Skelton}, \citenamefont
  {Frost}, \citenamefont {Terban}, \citenamefont {Zhai}, \citenamefont
  {Alatas}, \citenamefont {Owen}, \citenamefont {Walsh},\ and\ \citenamefont
  {Billinge}}]{beecher2016}%
  \BibitemOpen
  \bibfield  {author} {\bibinfo {author} {\bibfnamefont {A.~N.}\ \bibnamefont
  {Beecher}}, \bibinfo {author} {\bibfnamefont {O.~E.}\ \bibnamefont
  {Semonin}}, \bibinfo {author} {\bibfnamefont {J.~M.}\ \bibnamefont
  {Skelton}}, \bibinfo {author} {\bibfnamefont {J.~M.}\ \bibnamefont {Frost}},
  \bibinfo {author} {\bibfnamefont {M.~W.}\ \bibnamefont {Terban}}, \bibinfo
  {author} {\bibfnamefont {H.}~\bibnamefont {Zhai}}, \bibinfo {author}
  {\bibfnamefont {A.}~\bibnamefont {Alatas}}, \bibinfo {author} {\bibfnamefont
  {J.~S.}\ \bibnamefont {Owen}}, \bibinfo {author} {\bibfnamefont
  {A.}~\bibnamefont {Walsh}}, \ and\ \bibinfo {author} {\bibfnamefont {S.~J.}\
  \bibnamefont {Billinge}},\ }\href@noop {} {\bibfield  {journal} {\bibinfo
  {journal} {ACS energy lett.}\ }\textbf {\bibinfo {volume} {1}},\ \bibinfo
  {pages} {880} (\bibinfo {year} {2016})}\BibitemShut {NoStop}%
\bibitem [{\citenamefont {Adams}\ and\ \citenamefont
  {Passerone}(2016)}]{adams2016}%
  \BibitemOpen
  \bibfield  {author} {\bibinfo {author} {\bibfnamefont {D.~J.}\ \bibnamefont
  {Adams}}\ and\ \bibinfo {author} {\bibfnamefont {D.}~\bibnamefont
  {Passerone}},\ }\href {\doibase 10.1088/0953-8984/28/30/305401} {\bibfield
  {journal} {\bibinfo  {journal} {J. Phys. Cond. Matt.}\ }\textbf {\bibinfo
  {volume} {28}},\ \bibinfo {pages} {305401} (\bibinfo {year}
  {2016})}\BibitemShut {NoStop}%
\bibitem [{\citenamefont {Fujii}\ \emph {et~al.}(1974)\citenamefont {Fujii},
  \citenamefont {Hoshino}, \citenamefont {Yamada},\ and\ \citenamefont
  {Shirane}}]{fujii1974}%
  \BibitemOpen
  \bibfield  {author} {\bibinfo {author} {\bibfnamefont {Y.}~\bibnamefont
  {Fujii}}, \bibinfo {author} {\bibfnamefont {S.}~\bibnamefont {Hoshino}},
  \bibinfo {author} {\bibfnamefont {Y.}~\bibnamefont {Yamada}}, \ and\ \bibinfo
  {author} {\bibfnamefont {G.}~\bibnamefont {Shirane}},\ }\href@noop {}
  {\bibfield  {journal} {\bibinfo  {journal} {Phys. Rev. B}\ }\textbf {\bibinfo
  {volume} {9}},\ \bibinfo {pages} {4549} (\bibinfo {year} {1974})}\BibitemShut
  {NoStop}%
\bibitem [{\citenamefont {Chi}\ \emph {et~al.}(2005)\citenamefont {Chi},
  \citenamefont {Swainson}, \citenamefont {Cranswick}, \citenamefont {Her},
  \citenamefont {Stephens},\ and\ \citenamefont {Knop}}]{chi2005}%
  \BibitemOpen
  \bibfield  {author} {\bibinfo {author} {\bibfnamefont {L.}~\bibnamefont
  {Chi}}, \bibinfo {author} {\bibfnamefont {I.}~\bibnamefont {Swainson}},
  \bibinfo {author} {\bibfnamefont {L.}~\bibnamefont {Cranswick}}, \bibinfo
  {author} {\bibfnamefont {J.-H.}\ \bibnamefont {Her}}, \bibinfo {author}
  {\bibfnamefont {P.}~\bibnamefont {Stephens}}, \ and\ \bibinfo {author}
  {\bibfnamefont {O.}~\bibnamefont {Knop}},\ }\href@noop {} {\bibfield
  {journal} {\bibinfo  {journal} {J. Solid State Chem.}\ }\textbf {\bibinfo
  {volume} {178}},\ \bibinfo {pages} {1376} (\bibinfo {year}
  {2005})}\BibitemShut {NoStop}%
\bibitem [{\citenamefont {Swainson}\ \emph {et~al.}(2003)\citenamefont
  {Swainson}, \citenamefont {Hammond}, \citenamefont {Soulli{\`e}re},
  \citenamefont {Knop},\ and\ \citenamefont {Massa}}]{swainson2003}%
  \BibitemOpen
  \bibfield  {author} {\bibinfo {author} {\bibfnamefont {I.}~\bibnamefont
  {Swainson}}, \bibinfo {author} {\bibfnamefont {R.}~\bibnamefont {Hammond}},
  \bibinfo {author} {\bibfnamefont {C.}~\bibnamefont {Soulli{\`e}re}}, \bibinfo
  {author} {\bibfnamefont {O.}~\bibnamefont {Knop}}, \ and\ \bibinfo {author}
  {\bibfnamefont {W.}~\bibnamefont {Massa}},\ }\href@noop {} {\bibfield
  {journal} {\bibinfo  {journal} {J. Solid State Chem.}\ }\textbf {\bibinfo
  {volume} {176}},\ \bibinfo {pages} {97} (\bibinfo {year} {2003})}\BibitemShut
  {NoStop}%
\bibitem [{\citenamefont {Swainson}\ \emph {et~al.}(2015)\citenamefont
  {Swainson}, \citenamefont {Stock}, \citenamefont {Parker}, \citenamefont
  {Van~Eijck}, \citenamefont {Russina},\ and\ \citenamefont
  {Taylor}}]{swainson2015}%
  \BibitemOpen
  \bibfield  {author} {\bibinfo {author} {\bibfnamefont {I.}~\bibnamefont
  {Swainson}}, \bibinfo {author} {\bibfnamefont {C.}~\bibnamefont {Stock}},
  \bibinfo {author} {\bibfnamefont {S.}~\bibnamefont {Parker}}, \bibinfo
  {author} {\bibfnamefont {L.}~\bibnamefont {Van~Eijck}}, \bibinfo {author}
  {\bibfnamefont {M.}~\bibnamefont {Russina}}, \ and\ \bibinfo {author}
  {\bibfnamefont {J.}~\bibnamefont {Taylor}},\ }\href@noop {} {\bibfield
  {journal} {\bibinfo  {journal} {Phys. Rev. B}\ }\textbf {\bibinfo {volume}
  {92}},\ \bibinfo {pages} {100303} (\bibinfo {year} {2015})}\BibitemShut
  {NoStop}%
\bibitem [{\citenamefont {Poglitsch}\ and\ \citenamefont
  {Weber}(1987)}]{poglitsch1987}%
  \BibitemOpen
  \bibfield  {author} {\bibinfo {author} {\bibfnamefont {A.}~\bibnamefont
  {Poglitsch}}\ and\ \bibinfo {author} {\bibfnamefont {D.}~\bibnamefont
  {Weber}},\ }\href@noop {} {\bibfield  {journal} {\bibinfo  {journal} {J.
  Chem. Phys.}\ }\textbf {\bibinfo {volume} {87}},\ \bibinfo {pages} {6373}
  (\bibinfo {year} {1987})}\BibitemShut {NoStop}%
\bibitem [{\citenamefont {Egger}\ \emph {et~al.}(2018)\citenamefont {Egger},
  \citenamefont {Bera}, \citenamefont {Cahen}, \citenamefont {Hodes},
  \citenamefont {Kirchartz}, \citenamefont {Kronik}, \citenamefont {Lovrincic},
  \citenamefont {Rappe}, \citenamefont {Reichman},\ and\ \citenamefont
  {Yaffe}}]{egger2018}%
  \BibitemOpen
  \bibfield  {author} {\bibinfo {author} {\bibfnamefont {D.~A.}\ \bibnamefont
  {Egger}}, \bibinfo {author} {\bibfnamefont {A.}~\bibnamefont {Bera}},
  \bibinfo {author} {\bibfnamefont {D.}~\bibnamefont {Cahen}}, \bibinfo
  {author} {\bibfnamefont {G.}~\bibnamefont {Hodes}}, \bibinfo {author}
  {\bibfnamefont {T.}~\bibnamefont {Kirchartz}}, \bibinfo {author}
  {\bibfnamefont {L.}~\bibnamefont {Kronik}}, \bibinfo {author} {\bibfnamefont
  {R.}~\bibnamefont {Lovrincic}}, \bibinfo {author} {\bibfnamefont {A.~M.}\
  \bibnamefont {Rappe}}, \bibinfo {author} {\bibfnamefont {D.~R.}\ \bibnamefont
  {Reichman}}, \ and\ \bibinfo {author} {\bibfnamefont {O.}~\bibnamefont
  {Yaffe}},\ }\href@noop {} {\bibfield  {journal} {\bibinfo  {journal} {Adv.
  Materials}\ }\textbf {\bibinfo {volume} {30}},\ \bibinfo {pages} {1800691}
  (\bibinfo {year} {2018})}\BibitemShut {NoStop}%
\bibitem [{\citenamefont {Whalley}\ \emph {et~al.}(2017)\citenamefont
  {Whalley}, \citenamefont {Frost}, \citenamefont {Jung},\ and\ \citenamefont
  {Walsh}}]{whalley2017}%
  \BibitemOpen
  \bibfield  {author} {\bibinfo {author} {\bibfnamefont {L.~D.}\ \bibnamefont
  {Whalley}}, \bibinfo {author} {\bibfnamefont {J.~M.}\ \bibnamefont {Frost}},
  \bibinfo {author} {\bibfnamefont {Y.-K.}\ \bibnamefont {Jung}}, \ and\
  \bibinfo {author} {\bibfnamefont {A.}~\bibnamefont {Walsh}},\ }\href@noop {}
  {\bibfield  {journal} {\bibinfo  {journal} {J. Chem. Phys.}\ }\textbf
  {\bibinfo {volume} {146}},\ \bibinfo {pages} {220901} (\bibinfo {year}
  {2017})}\BibitemShut {NoStop}%
\bibitem [{\citenamefont {Onoda-Yamamuro}\ \emph {et~al.}(1992)\citenamefont
  {Onoda-Yamamuro}, \citenamefont {Yamamuro}, \citenamefont {Matsuo},\ and\
  \citenamefont {Suga}}]{onoda1992}%
  \BibitemOpen
  \bibfield  {author} {\bibinfo {author} {\bibfnamefont {N.}~\bibnamefont
  {Onoda-Yamamuro}}, \bibinfo {author} {\bibfnamefont {O.}~\bibnamefont
  {Yamamuro}}, \bibinfo {author} {\bibfnamefont {T.}~\bibnamefont {Matsuo}}, \
  and\ \bibinfo {author} {\bibfnamefont {H.}~\bibnamefont {Suga}},\ }\href@noop
  {} {\bibfield  {journal} {\bibinfo  {journal} {J. Phys. Chem. Solids}\
  }\textbf {\bibinfo {volume} {53}},\ \bibinfo {pages} {277} (\bibinfo {year}
  {1992})}\BibitemShut {NoStop}%
\bibitem [{\citenamefont {Lin}\ \emph {et~al.}(2021)\citenamefont {Lin},
  \citenamefont {Zhang}, \citenamefont {Gao}, \citenamefont {Steele},
  \citenamefont {Louisia}, \citenamefont {Yu}, \citenamefont {Quan},
  \citenamefont {Lin}, \citenamefont {Limmer},\ and\ \citenamefont
  {Yang}}]{lin2021}%
  \BibitemOpen
  \bibfield  {author} {\bibinfo {author} {\bibfnamefont {Z.}~\bibnamefont
  {Lin}}, \bibinfo {author} {\bibfnamefont {Y.}~\bibnamefont {Zhang}}, \bibinfo
  {author} {\bibfnamefont {M.}~\bibnamefont {Gao}}, \bibinfo {author}
  {\bibfnamefont {J.~A.}\ \bibnamefont {Steele}}, \bibinfo {author}
  {\bibfnamefont {S.}~\bibnamefont {Louisia}}, \bibinfo {author} {\bibfnamefont
  {S.}~\bibnamefont {Yu}}, \bibinfo {author} {\bibfnamefont {L.~N.}\
  \bibnamefont {Quan}}, \bibinfo {author} {\bibfnamefont {C.-K.}\ \bibnamefont
  {Lin}}, \bibinfo {author} {\bibfnamefont {D.~T.}\ \bibnamefont {Limmer}}, \
  and\ \bibinfo {author} {\bibfnamefont {P.}~\bibnamefont {Yang}},\ }\href@noop
  {} {\bibfield  {journal} {\bibinfo  {journal} {Matter}\ }\textbf {\bibinfo
  {volume} {4}},\ \bibinfo {pages} {2392} (\bibinfo {year} {2021})}\BibitemShut
  {NoStop}%
\bibitem [{\citenamefont {Ugur}\ \emph {et~al.}(2020)\citenamefont {Ugur},
  \citenamefont {Alarousu}, \citenamefont {Khan}, \citenamefont {Vlk},
  \citenamefont {Aydin}, \citenamefont {De~Bastiani}, \citenamefont {Balawi},
  \citenamefont {Gonzalez~Lopez}, \citenamefont {Ledinsk{\`y}}, \citenamefont
  {De~Wolf} \emph {et~al.}}]{ugur2020}%
  \BibitemOpen
  \bibfield  {author} {\bibinfo {author} {\bibfnamefont {E.}~\bibnamefont
  {Ugur}}, \bibinfo {author} {\bibfnamefont {E.}~\bibnamefont {Alarousu}},
  \bibinfo {author} {\bibfnamefont {J.~I.}\ \bibnamefont {Khan}}, \bibinfo
  {author} {\bibfnamefont {A.}~\bibnamefont {Vlk}}, \bibinfo {author}
  {\bibfnamefont {E.}~\bibnamefont {Aydin}}, \bibinfo {author} {\bibfnamefont
  {M.}~\bibnamefont {De~Bastiani}}, \bibinfo {author} {\bibfnamefont {A.~H.}\
  \bibnamefont {Balawi}}, \bibinfo {author} {\bibfnamefont {S.~P.}\
  \bibnamefont {Gonzalez~Lopez}}, \bibinfo {author} {\bibfnamefont
  {M.}~\bibnamefont {Ledinsk{\`y}}}, \bibinfo {author} {\bibfnamefont
  {S.}~\bibnamefont {De~Wolf}},  \emph {et~al.},\ }\href@noop {} {\bibfield
  {journal} {\bibinfo  {journal} {Solar RRL}\ }\textbf {\bibinfo {volume}
  {4}},\ \bibinfo {pages} {2000382} (\bibinfo {year} {2020})}\BibitemShut
  {NoStop}%
\bibitem [{\citenamefont {Glazer}(1972)}]{glazer1972}%
  \BibitemOpen
  \bibfield  {author} {\bibinfo {author} {\bibfnamefont {A.}~\bibnamefont
  {Glazer}},\ }\href@noop {} {\bibfield  {journal} {\bibinfo  {journal} {Acta
  Cryst. B}\ }\textbf {\bibinfo {volume} {28}},\ \bibinfo {pages} {3384}
  (\bibinfo {year} {1972})}\BibitemShut {NoStop}%
\bibitem [{\citenamefont {Marronnier}\ \emph {et~al.}(2017)\citenamefont
  {Marronnier}, \citenamefont {Lee}, \citenamefont {Geffroy}, \citenamefont
  {Even}, \citenamefont {Bonnassieux},\ and\ \citenamefont
  {Roma}}]{marronnier2017}%
  \BibitemOpen
  \bibfield  {author} {\bibinfo {author} {\bibfnamefont {A.}~\bibnamefont
  {Marronnier}}, \bibinfo {author} {\bibfnamefont {H.}~\bibnamefont {Lee}},
  \bibinfo {author} {\bibfnamefont {B.}~\bibnamefont {Geffroy}}, \bibinfo
  {author} {\bibfnamefont {J.}~\bibnamefont {Even}}, \bibinfo {author}
  {\bibfnamefont {Y.}~\bibnamefont {Bonnassieux}}, \ and\ \bibinfo {author}
  {\bibfnamefont {G.}~\bibnamefont {Roma}},\ }\href@noop {} {\bibfield
  {journal} {\bibinfo  {journal} {J. Phys. Chem. Lett.}\ }\textbf {\bibinfo
  {volume} {8}},\ \bibinfo {pages} {2659} (\bibinfo {year} {2017})}\BibitemShut
  {NoStop}%
\bibitem [{\citenamefont {Stokes}\ and\ \citenamefont
  {Hatch}(2005)}]{stokes2005}%
  \BibitemOpen
  \bibfield  {author} {\bibinfo {author} {\bibfnamefont {H.~T.}\ \bibnamefont
  {Stokes}}\ and\ \bibinfo {author} {\bibfnamefont {D.~M.}\ \bibnamefont
  {Hatch}},\ }\href@noop {} {\bibfield  {journal} {\bibinfo  {journal} {J. App.
  Cryst.}\ }\textbf {\bibinfo {volume} {38}},\ \bibinfo {pages} {237} (\bibinfo
  {year} {2005})}\BibitemShut {NoStop}%
\bibitem [{\citenamefont {Howard}\ and\ \citenamefont
  {Stokes}(1998)}]{howard1998}%
  \BibitemOpen
  \bibfield  {author} {\bibinfo {author} {\bibfnamefont {C.~J.}\ \bibnamefont
  {Howard}}\ and\ \bibinfo {author} {\bibfnamefont {H.~T.}\ \bibnamefont
  {Stokes}},\ }\href@noop {} {\bibfield  {journal} {\bibinfo  {journal} {Acta
  Cryst. B}\ }\textbf {\bibinfo {volume} {54}},\ \bibinfo {pages} {782}
  (\bibinfo {year} {1998})}\BibitemShut {NoStop}%
\bibitem [{\citenamefont {Jacobsson}\ \emph {et~al.}(2015)\citenamefont
  {Jacobsson}, \citenamefont {Schwan}, \citenamefont {Ottosson}, \citenamefont
  {Hagfeldt},\ and\ \citenamefont {Edvinsson}}]{jacobsson2015}%
  \BibitemOpen
  \bibfield  {author} {\bibinfo {author} {\bibfnamefont {T.~J.}\ \bibnamefont
  {Jacobsson}}, \bibinfo {author} {\bibfnamefont {L.~J.}\ \bibnamefont
  {Schwan}}, \bibinfo {author} {\bibfnamefont {M.}~\bibnamefont {Ottosson}},
  \bibinfo {author} {\bibfnamefont {A.}~\bibnamefont {Hagfeldt}}, \ and\
  \bibinfo {author} {\bibfnamefont {T.}~\bibnamefont {Edvinsson}},\ }\href@noop
  {} {\bibfield  {journal} {\bibinfo  {journal} {Inorg. Chem.}\ }\textbf
  {\bibinfo {volume} {54}},\ \bibinfo {pages} {10678} (\bibinfo {year}
  {2015})}\BibitemShut {NoStop}%
\bibitem [{\citenamefont {Brivio}\ \emph {et~al.}(2015)\citenamefont {Brivio},
  \citenamefont {Frost}, \citenamefont {Skelton}, \citenamefont {Jackson},
  \citenamefont {Weber}, \citenamefont {Weller}, \citenamefont {Goni},
  \citenamefont {Leguy}, \citenamefont {Barnes},\ and\ \citenamefont
  {Walsh}}]{brivio2015}%
  \BibitemOpen
  \bibfield  {author} {\bibinfo {author} {\bibfnamefont {F.}~\bibnamefont
  {Brivio}}, \bibinfo {author} {\bibfnamefont {J.~M.}\ \bibnamefont {Frost}},
  \bibinfo {author} {\bibfnamefont {J.~M.}\ \bibnamefont {Skelton}}, \bibinfo
  {author} {\bibfnamefont {A.~J.}\ \bibnamefont {Jackson}}, \bibinfo {author}
  {\bibfnamefont {O.~J.}\ \bibnamefont {Weber}}, \bibinfo {author}
  {\bibfnamefont {M.~T.}\ \bibnamefont {Weller}}, \bibinfo {author}
  {\bibfnamefont {A.~R.}\ \bibnamefont {Goni}}, \bibinfo {author}
  {\bibfnamefont {A.~M.}\ \bibnamefont {Leguy}}, \bibinfo {author}
  {\bibfnamefont {P.~R.}\ \bibnamefont {Barnes}}, \ and\ \bibinfo {author}
  {\bibfnamefont {A.}~\bibnamefont {Walsh}},\ }\href@noop {} {\bibfield
  {journal} {\bibinfo  {journal} {Phys. Rev. B}\ }\textbf {\bibinfo {volume}
  {92}},\ \bibinfo {pages} {144308} (\bibinfo {year} {2015})}\BibitemShut
  {NoStop}%
\bibitem [{\citenamefont {Tyson}\ \emph {et~al.}(2017)\citenamefont {Tyson},
  \citenamefont {Gao}, \citenamefont {Chen}, \citenamefont {Ghose},\ and\
  \citenamefont {Yan}}]{tyson2017}%
  \BibitemOpen
  \bibfield  {author} {\bibinfo {author} {\bibfnamefont {T.}~\bibnamefont
  {Tyson}}, \bibinfo {author} {\bibfnamefont {W.}~\bibnamefont {Gao}}, \bibinfo
  {author} {\bibfnamefont {Y.-S.}\ \bibnamefont {Chen}}, \bibinfo {author}
  {\bibfnamefont {S.}~\bibnamefont {Ghose}}, \ and\ \bibinfo {author}
  {\bibfnamefont {Y.}~\bibnamefont {Yan}},\ }\href@noop {} {\bibfield
  {journal} {\bibinfo  {journal} {Scientific reports}\ }\textbf {\bibinfo
  {volume} {7}},\ \bibinfo {pages} {9401} (\bibinfo {year} {2017})}\BibitemShut
  {NoStop}%
\bibitem [{\citenamefont {Sakakura}\ \emph {et~al.}(2011)\citenamefont
  {Sakakura}, \citenamefont {Wang}, \citenamefont {Ishizawa}, \citenamefont
  {Inagaki},\ and\ \citenamefont {Kakimoto}}]{sakakura2011}%
  \BibitemOpen
  \bibfield  {author} {\bibinfo {author} {\bibfnamefont {T.}~\bibnamefont
  {Sakakura}}, \bibinfo {author} {\bibfnamefont {J.}~\bibnamefont {Wang}},
  \bibinfo {author} {\bibfnamefont {N.}~\bibnamefont {Ishizawa}}, \bibinfo
  {author} {\bibfnamefont {Y.}~\bibnamefont {Inagaki}}, \ and\ \bibinfo
  {author} {\bibfnamefont {K.}~\bibnamefont {Kakimoto}},\ }in\ \href@noop {}
  {\emph {\bibinfo {booktitle} {IOP Conference Series: Materials Science and
  Engineering}}},\ Vol.~\bibinfo {volume} {18}\ (\bibinfo {organization} {IOP
  Publishing},\ \bibinfo {year} {2011})\ p.\ \bibinfo {pages}
  {022006}\BibitemShut {NoStop}%
\bibitem [{\citenamefont {Kn{\'\i}{\v{z}}ek}\ \emph {et~al.}(2004)\citenamefont
  {Kn{\'\i}{\v{z}}ek}, \citenamefont {Hejtmanek}, \citenamefont {Jirak},
  \citenamefont {Martin}, \citenamefont {Hervieu}, \citenamefont {Raveau},
  \citenamefont {Andr{\'e}},\ and\ \citenamefont {Bouree}}]{knivzek2004}%
  \BibitemOpen
  \bibfield  {author} {\bibinfo {author} {\bibfnamefont {K.}~\bibnamefont
  {Kn{\'\i}{\v{z}}ek}}, \bibinfo {author} {\bibfnamefont {J.}~\bibnamefont
  {Hejtmanek}}, \bibinfo {author} {\bibfnamefont {Z.}~\bibnamefont {Jirak}},
  \bibinfo {author} {\bibfnamefont {C.}~\bibnamefont {Martin}}, \bibinfo
  {author} {\bibfnamefont {M.}~\bibnamefont {Hervieu}}, \bibinfo {author}
  {\bibfnamefont {B.}~\bibnamefont {Raveau}}, \bibinfo {author} {\bibfnamefont
  {G.}~\bibnamefont {Andr{\'e}}}, \ and\ \bibinfo {author} {\bibfnamefont
  {F.}~\bibnamefont {Bouree}},\ }\href@noop {} {\bibfield  {journal} {\bibinfo
  {journal} {Chem. Materials}\ }\textbf {\bibinfo {volume} {16}},\ \bibinfo
  {pages} {1104} (\bibinfo {year} {2004})}\BibitemShut {NoStop}%
\bibitem [{\citenamefont {Heyraud}\ \emph {et~al.}(2013)\citenamefont
  {Heyraud}, \citenamefont {Blanchard}, \citenamefont {Liu}, \citenamefont
  {Zhou}, \citenamefont {Kennedy}, \citenamefont {Brand}, \citenamefont
  {Tadich},\ and\ \citenamefont {Hester}}]{heyraud2013}%
  \BibitemOpen
  \bibfield  {author} {\bibinfo {author} {\bibfnamefont {S.}~\bibnamefont
  {Heyraud}}, \bibinfo {author} {\bibfnamefont {P.~E.}\ \bibnamefont
  {Blanchard}}, \bibinfo {author} {\bibfnamefont {S.}~\bibnamefont {Liu}},
  \bibinfo {author} {\bibfnamefont {Q.}~\bibnamefont {Zhou}}, \bibinfo {author}
  {\bibfnamefont {B.~J.}\ \bibnamefont {Kennedy}}, \bibinfo {author}
  {\bibfnamefont {H.~E.}\ \bibnamefont {Brand}}, \bibinfo {author}
  {\bibfnamefont {A.}~\bibnamefont {Tadich}}, \ and\ \bibinfo {author}
  {\bibfnamefont {J.~R.}\ \bibnamefont {Hester}},\ }\href@noop {} {\bibfield
  {journal} {\bibinfo  {journal} {J. Phys. Cond. Matt.}\ }\textbf {\bibinfo
  {volume} {25}},\ \bibinfo {pages} {335401} (\bibinfo {year}
  {2013})}\BibitemShut {NoStop}%
\bibitem [{\citenamefont {Kassan-Ogly}\ and\ \citenamefont
  {Naish}(1986)}]{kassan1986}%
  \BibitemOpen
  \bibfield  {author} {\bibinfo {author} {\bibfnamefont {F.}~\bibnamefont
  {Kassan-Ogly}}\ and\ \bibinfo {author} {\bibfnamefont {V.}~\bibnamefont
  {Naish}},\ }\href@noop {} {\bibfield  {journal} {\bibinfo  {journal} {Acta
  Cryst. B}\ }\textbf {\bibinfo {volume} {42}},\ \bibinfo {pages} {325}
  (\bibinfo {year} {1986})}\BibitemShut {NoStop}%
\bibitem [{Note1()}]{Note1}%
  \BibitemOpen
  \bibinfo {note} {This can be described in terms of Fermi's golden rule for
  the transition rate $\Gamma _{i\rightarrow f}= \protect \frac {2\pi }{\hbar }
  \left | \langle f | V | i \rangle \right | ^2 \rho $, where $f$ and $i$ are
  final and the initial state respectively, $\rho $ the density of final states
  and $\langle f|V|i\rangle $ matrix element connecting the two
  states.}\BibitemShut {Stop}%
\bibitem [{\citenamefont {Zhang}\ \emph {et~al.}(2015)\citenamefont {Zhang},
  \citenamefont {Pa{\'{s}}ciak}, \citenamefont {Glazer}, \citenamefont
  {Hlinka}, \citenamefont {Gutmann}, \citenamefont {Sparkes}, \citenamefont
  {Welberry}, \citenamefont {Majchrowski}, \citenamefont {Roleder},
  \citenamefont {Xie},\ and\ \citenamefont {Ye}}]{zhang2015}%
  \BibitemOpen
  \bibfield  {author} {\bibinfo {author} {\bibfnamefont {N.}~\bibnamefont
  {Zhang}}, \bibinfo {author} {\bibfnamefont {M.}~\bibnamefont
  {Pa{\'{s}}ciak}}, \bibinfo {author} {\bibfnamefont {A.~M.}\ \bibnamefont
  {Glazer}}, \bibinfo {author} {\bibfnamefont {J.}~\bibnamefont {Hlinka}},
  \bibinfo {author} {\bibfnamefont {M.}~\bibnamefont {Gutmann}}, \bibinfo
  {author} {\bibfnamefont {H.~A.}\ \bibnamefont {Sparkes}}, \bibinfo {author}
  {\bibfnamefont {T.~R.}\ \bibnamefont {Welberry}}, \bibinfo {author}
  {\bibfnamefont {A.}~\bibnamefont {Majchrowski}}, \bibinfo {author}
  {\bibfnamefont {K.}~\bibnamefont {Roleder}}, \bibinfo {author} {\bibfnamefont
  {Y.}~\bibnamefont {Xie}}, \ and\ \bibinfo {author} {\bibfnamefont {Z.-G.}\
  \bibnamefont {Ye}},\ }\href {\doibase 10.1107/S1600576715017069} {\bibfield
  {journal} {\bibinfo  {journal} {J. Applied Crystallography}\ }\textbf
  {\bibinfo {volume} {48}},\ \bibinfo {pages} {1637} (\bibinfo {year}
  {2015})}\BibitemShut {NoStop}%
\bibitem [{Note2()}]{Note2}%
  \BibitemOpen
  \bibinfo {note} {At room temperature a centrosymmetric structure (space group
  $Pbam$) is observed, resulting from antiparallel displacements of the cations
  on the (110) planes and oxygen octahedral tilts of type $a^-a^-c^0$ \cite
  {glazer1993}.}\BibitemShut {Stop}%
\bibitem [{\citenamefont {Aleksandrov}(1976)}]{aleksandrov1976}%
  \BibitemOpen
  \bibfield  {author} {\bibinfo {author} {\bibfnamefont {K.}~\bibnamefont
  {Aleksandrov}},\ }\href@noop {} {\bibfield  {journal} {\bibinfo  {journal}
  {Ferroelectrics}\ }\textbf {\bibinfo {volume} {24}},\ \bibinfo {pages} {801}
  (\bibinfo {year} {1976})}\BibitemShut {NoStop}%
\bibitem [{\citenamefont {Hossain}\ \emph {et~al.}(2018)\citenamefont
  {Hossain}, \citenamefont {Bandyopadhyay},\ and\ \citenamefont
  {Roy}}]{hossain2018}%
  \BibitemOpen
  \bibfield  {author} {\bibinfo {author} {\bibfnamefont {A.}~\bibnamefont
  {Hossain}}, \bibinfo {author} {\bibfnamefont {P.}~\bibnamefont
  {Bandyopadhyay}}, \ and\ \bibinfo {author} {\bibfnamefont {S.}~\bibnamefont
  {Roy}},\ }\href@noop {} {\bibfield  {journal} {\bibinfo  {journal} {J. Alloys
  and Compounds}\ }\textbf {\bibinfo {volume} {740}},\ \bibinfo {pages} {414 }
  (\bibinfo {year} {2018})}\BibitemShut {NoStop}%
\bibitem [{\citenamefont {Bristowe}\ \emph {et~al.}(2015)\citenamefont
  {Bristowe}, \citenamefont {Varignon}, \citenamefont {Fontaine}, \citenamefont
  {Bousquet},\ and\ \citenamefont {Ghosez}}]{bristowe2015}%
  \BibitemOpen
  \bibfield  {author} {\bibinfo {author} {\bibfnamefont {N.}~\bibnamefont
  {Bristowe}}, \bibinfo {author} {\bibfnamefont {J.}~\bibnamefont {Varignon}},
  \bibinfo {author} {\bibfnamefont {D.}~\bibnamefont {Fontaine}}, \bibinfo
  {author} {\bibfnamefont {E.}~\bibnamefont {Bousquet}}, \ and\ \bibinfo
  {author} {\bibfnamefont {P.}~\bibnamefont {Ghosez}},\ }\href@noop {}
  {\bibfield  {journal} {\bibinfo  {journal} {Nature communications}\ }\textbf
  {\bibinfo {volume} {6}},\ \bibinfo {pages} {6677} (\bibinfo {year}
  {2015})}\BibitemShut {NoStop}%
\bibitem [{\citenamefont {Buttner}\ and\ \citenamefont
  {Maslen}(1992)}]{buttner1992}%
  \BibitemOpen
  \bibfield  {author} {\bibinfo {author} {\bibfnamefont {R.}~\bibnamefont
  {Buttner}}\ and\ \bibinfo {author} {\bibfnamefont {E.}~\bibnamefont
  {Maslen}},\ }\href@noop {} {\bibfield  {journal} {\bibinfo  {journal} {Acta
  Cryst. B}\ }\textbf {\bibinfo {volume} {48}},\ \bibinfo {pages} {764}
  (\bibinfo {year} {1992})}\BibitemShut {NoStop}%
\bibitem [{\citenamefont {Edwards}\ and\ \citenamefont
  {Peacock}(1959)}]{edwards1959}%
  \BibitemOpen
  \bibfield  {author} {\bibinfo {author} {\bibfnamefont {A.}~\bibnamefont
  {Edwards}}\ and\ \bibinfo {author} {\bibfnamefont {R.}~\bibnamefont
  {Peacock}},\ }\href@noop {} {\bibfield  {journal} {\bibinfo  {journal} {J.
  Chem. Soc.}\ ,\ \bibinfo {pages} {4126}} (\bibinfo {year}
  {1959})}\BibitemShut {NoStop}%
\bibitem [{\citenamefont {Tanaka}\ \emph {et~al.}(1993)\citenamefont {Tanaka},
  \citenamefont {Shishido}, \citenamefont {Horiuchi}, \citenamefont {Toyota},
  \citenamefont {Shindo},\ and\ \citenamefont {Fukuda}}]{tanaka1993}%
  \BibitemOpen
  \bibfield  {author} {\bibinfo {author} {\bibfnamefont {M.}~\bibnamefont
  {Tanaka}}, \bibinfo {author} {\bibfnamefont {T.}~\bibnamefont {Shishido}},
  \bibinfo {author} {\bibfnamefont {H.}~\bibnamefont {Horiuchi}}, \bibinfo
  {author} {\bibfnamefont {N.}~\bibnamefont {Toyota}}, \bibinfo {author}
  {\bibfnamefont {D.}~\bibnamefont {Shindo}}, \ and\ \bibinfo {author}
  {\bibfnamefont {T.}~\bibnamefont {Fukuda}},\ }\href@noop {} {\bibfield
  {journal} {\bibinfo  {journal} {J. Alloys and Compounds}\ }\textbf {\bibinfo
  {volume} {192}},\ \bibinfo {pages} {87} (\bibinfo {year} {1993})}\BibitemShut
  {NoStop}%
\bibitem [{\citenamefont {R{\"u}dorff}\ \emph {et~al.}(1963)\citenamefont
  {R{\"u}dorff}, \citenamefont {Lincke},\ and\ \citenamefont
  {Babel}}]{rudorff1963}%
  \BibitemOpen
  \bibfield  {author} {\bibinfo {author} {\bibfnamefont {W.}~\bibnamefont
  {R{\"u}dorff}}, \bibinfo {author} {\bibfnamefont {G.}~\bibnamefont {Lincke}},
  \ and\ \bibinfo {author} {\bibfnamefont {D.}~\bibnamefont {Babel}},\
  }\href@noop {} {\bibfield  {journal} {\bibinfo  {journal} {Z. f{\"u}r
  anorganische und allgemeine Chemie}\ }\textbf {\bibinfo {volume} {320}},\
  \bibinfo {pages} {150} (\bibinfo {year} {1963})}\BibitemShut {NoStop}%
\bibitem [{\citenamefont {Diodati}\ \emph {et~al.}(2012)\citenamefont
  {Diodati}, \citenamefont {Nodari}, \citenamefont {Natile}, \citenamefont
  {Russo}, \citenamefont {Tondello}, \citenamefont {Lutterotti},\ and\
  \citenamefont {Gross}}]{diodati2012}%
  \BibitemOpen
  \bibfield  {author} {\bibinfo {author} {\bibfnamefont {S.}~\bibnamefont
  {Diodati}}, \bibinfo {author} {\bibfnamefont {L.}~\bibnamefont {Nodari}},
  \bibinfo {author} {\bibfnamefont {M.}~\bibnamefont {Natile}}, \bibinfo
  {author} {\bibfnamefont {U.}~\bibnamefont {Russo}}, \bibinfo {author}
  {\bibfnamefont {E.}~\bibnamefont {Tondello}}, \bibinfo {author}
  {\bibfnamefont {L.}~\bibnamefont {Lutterotti}}, \ and\ \bibinfo {author}
  {\bibfnamefont {S.}~\bibnamefont {Gross}},\ }\href@noop {} {\bibfield
  {journal} {\bibinfo  {journal} {Dalton Transactions}\ }\textbf {\bibinfo
  {volume} {41}},\ \bibinfo {pages} {5517} (\bibinfo {year}
  {2012})}\BibitemShut {NoStop}%
\bibitem [{\citenamefont {Matsushita}\ \emph {et~al.}(2007)\citenamefont
  {Matsushita}, \citenamefont {Ahsbahs}, \citenamefont {Hafner},\ and\
  \citenamefont {Kojima}}]{matsushita2007}%
  \BibitemOpen
  \bibfield  {author} {\bibinfo {author} {\bibfnamefont {N.}~\bibnamefont
  {Matsushita}}, \bibinfo {author} {\bibfnamefont {H.}~\bibnamefont {Ahsbahs}},
  \bibinfo {author} {\bibfnamefont {S.~S.}\ \bibnamefont {Hafner}}, \ and\
  \bibinfo {author} {\bibfnamefont {N.}~\bibnamefont {Kojima}},\ }\href@noop {}
  {\bibfield  {journal} {\bibinfo  {journal} {J. Solid State Chem.}\ }\textbf
  {\bibinfo {volume} {180}},\ \bibinfo {pages} {1353} (\bibinfo {year}
  {2007})}\BibitemShut {NoStop}%
\bibitem [{\citenamefont {Shishido}\ \emph {et~al.}(1997)\citenamefont
  {Shishido}, \citenamefont {Zheng}, \citenamefont {Saito}, \citenamefont
  {Horiuchi}, \citenamefont {Kudou}, \citenamefont {Okada},\ and\ \citenamefont
  {Fukuda}}]{shishido1997}%
  \BibitemOpen
  \bibfield  {author} {\bibinfo {author} {\bibfnamefont {T.}~\bibnamefont
  {Shishido}}, \bibinfo {author} {\bibfnamefont {Y.}~\bibnamefont {Zheng}},
  \bibinfo {author} {\bibfnamefont {A.}~\bibnamefont {Saito}}, \bibinfo
  {author} {\bibfnamefont {H.}~\bibnamefont {Horiuchi}}, \bibinfo {author}
  {\bibfnamefont {K.}~\bibnamefont {Kudou}}, \bibinfo {author} {\bibfnamefont
  {S.}~\bibnamefont {Okada}}, \ and\ \bibinfo {author} {\bibfnamefont
  {T.}~\bibnamefont {Fukuda}},\ }\href@noop {} {\bibfield  {journal} {\bibinfo
  {journal} {J. Alloys and Compounds}\ }\textbf {\bibinfo {volume} {260}},\
  \bibinfo {pages} {88} (\bibinfo {year} {1997})}\BibitemShut {NoStop}%
\bibitem [{\citenamefont {Hayward}\ and\ \citenamefont
  {Salje}(2002)}]{hayward2002}%
  \BibitemOpen
  \bibfield  {author} {\bibinfo {author} {\bibfnamefont {S.}~\bibnamefont
  {Hayward}}\ and\ \bibinfo {author} {\bibfnamefont {E.}~\bibnamefont
  {Salje}},\ }\href@noop {} {\bibfield  {journal} {\bibinfo  {journal} {J.
  Phys. Cond. Matt.}\ }\textbf {\bibinfo {volume} {14}},\ \bibinfo {pages}
  {L599} (\bibinfo {year} {2002})}\BibitemShut {NoStop}%
\bibitem [{\citenamefont {Sanz}\ \emph {et~al.}(2004)\citenamefont {Sanz},
  \citenamefont {Varez}, \citenamefont {Alonso},\ and\ \citenamefont
  {Fernandez}}]{sanz2004}%
  \BibitemOpen
  \bibfield  {author} {\bibinfo {author} {\bibfnamefont {J.}~\bibnamefont
  {Sanz}}, \bibinfo {author} {\bibfnamefont {A.}~\bibnamefont {Varez}},
  \bibinfo {author} {\bibfnamefont {J.~A.}\ \bibnamefont {Alonso}}, \ and\
  \bibinfo {author} {\bibfnamefont {M.~T.}\ \bibnamefont {Fernandez}},\
  }\href@noop {} {\bibfield  {journal} {\bibinfo  {journal} {J. Solid State
  Chem.}\ }\textbf {\bibinfo {volume} {177}},\ \bibinfo {pages} {1157}
  (\bibinfo {year} {2004})}\BibitemShut {NoStop}%
\bibitem [{\citenamefont {Zachariasen}(1928)}]{zachariasen1928}%
  \BibitemOpen
  \bibfield  {author} {\bibinfo {author} {\bibfnamefont {W.}~\bibnamefont
  {Zachariasen}},\ }\href@noop {} {\bibfield  {journal} {\bibinfo  {journal}
  {Matematisk Naturvidenskapelig Klasse}\ ,\ \bibinfo {pages} {1}} (\bibinfo
  {year} {1928})}\BibitemShut {NoStop}%
\bibitem [{\citenamefont {Sefat}\ \emph {et~al.}(2005)\citenamefont {Sefat},
  \citenamefont {Amow}, \citenamefont {Wu}, \citenamefont {Botton},\ and\
  \citenamefont {Greedan}}]{sefat2005}%
  \BibitemOpen
  \bibfield  {author} {\bibinfo {author} {\bibfnamefont {A.~S.}\ \bibnamefont
  {Sefat}}, \bibinfo {author} {\bibfnamefont {G.}~\bibnamefont {Amow}},
  \bibinfo {author} {\bibfnamefont {M.-Y.}\ \bibnamefont {Wu}}, \bibinfo
  {author} {\bibfnamefont {G.~A.}\ \bibnamefont {Botton}}, \ and\ \bibinfo
  {author} {\bibfnamefont {J.}~\bibnamefont {Greedan}},\ }\href@noop {}
  {\bibfield  {journal} {\bibinfo  {journal} {J. Solid State Chem.}\ }\textbf
  {\bibinfo {volume} {178}},\ \bibinfo {pages} {1008} (\bibinfo {year}
  {2005})}\BibitemShut {NoStop}%
\bibitem [{\citenamefont {Cole}\ and\ \citenamefont
  {Espenschied}(1937)}]{cole1937}%
  \BibitemOpen
  \bibfield  {author} {\bibinfo {author} {\bibfnamefont {S.~S.}\ \bibnamefont
  {Cole}}\ and\ \bibinfo {author} {\bibfnamefont {H.}~\bibnamefont
  {Espenschied}},\ }\href@noop {} {\bibfield  {journal} {\bibinfo  {journal}
  {J. phys. chem.}\ }\textbf {\bibinfo {volume} {41}},\ \bibinfo {pages} {445}
  (\bibinfo {year} {1937})}\BibitemShut {NoStop}%
\bibitem [{\citenamefont {Zhu}\ \emph {et~al.}(2011)\citenamefont {Zhu},
  \citenamefont {Xu}, \citenamefont {Zhang}, \citenamefont {Jin}, \citenamefont
  {Wang},\ and\ \citenamefont {Zhao}}]{zhu2011}%
  \BibitemOpen
  \bibfield  {author} {\bibinfo {author} {\bibfnamefont {J.}~\bibnamefont
  {Zhu}}, \bibinfo {author} {\bibfnamefont {H.}~\bibnamefont {Xu}}, \bibinfo
  {author} {\bibfnamefont {J.}~\bibnamefont {Zhang}}, \bibinfo {author}
  {\bibfnamefont {C.}~\bibnamefont {Jin}}, \bibinfo {author} {\bibfnamefont
  {L.}~\bibnamefont {Wang}}, \ and\ \bibinfo {author} {\bibfnamefont
  {Y.}~\bibnamefont {Zhao}},\ }\href@noop {} {\bibfield  {journal} {\bibinfo
  {journal} {J. Appl. Phys.}\ }\textbf {\bibinfo {volume} {110}},\ \bibinfo
  {pages} {084103} (\bibinfo {year} {2011})}\BibitemShut {NoStop}%
\bibitem [{\citenamefont {Solovev}\ \emph {et~al.}(1961)\citenamefont
  {Solovev}, \citenamefont {Venevtsev},\ and\ \citenamefont
  {Zhanov}}]{solov1961}%
  \BibitemOpen
  \bibfield  {author} {\bibinfo {author} {\bibfnamefont {S.}~\bibnamefont
  {Solovev}}, \bibinfo {author} {\bibfnamefont {Y.~N.}\ \bibnamefont
  {Venevtsev}}, \ and\ \bibinfo {author} {\bibfnamefont {G.}~\bibnamefont
  {Zhanov}},\ }\href@noop {} {\bibfield  {journal} {\bibinfo  {journal} {Soviet
  Physics Crystallography}\ }\textbf {\bibinfo {volume} {6}},\ \bibinfo {pages}
  {171} (\bibinfo {year} {1961})}\BibitemShut {NoStop}%
\bibitem [{\citenamefont {Zaslavskii}\ and\ \citenamefont
  {Bryzhina}(1963)}]{zaslavskii1963}%
  \BibitemOpen
  \bibfield  {author} {\bibinfo {author} {\bibfnamefont {A.}~\bibnamefont
  {Zaslavskii}}\ and\ \bibinfo {author} {\bibfnamefont {M.}~\bibnamefont
  {Bryzhina}},\ }\href@noop {} {\bibfield  {journal} {\bibinfo  {journal} {Sov.
  Phys. Crystallogr-}\ }\textbf {\bibinfo {volume} {7}},\ \bibinfo {pages}
  {577} (\bibinfo {year} {1963})}\BibitemShut {NoStop}%
\bibitem [{\citenamefont {Ruggiero}\ and\ \citenamefont
  {Ferro}(1954)}]{ruggiero1954}%
  \BibitemOpen
  \bibfield  {author} {\bibinfo {author} {\bibfnamefont {A.}~\bibnamefont
  {Ruggiero}}\ and\ \bibinfo {author} {\bibfnamefont {R.}~\bibnamefont
  {Ferro}},\ }\href@noop {} {\bibfield  {journal} {\bibinfo  {journal} {Atti
  della Accademia Nazionale dei Lincei, Classe di Scienze Fisiche, Matematiche
  e Naturali}\ }\textbf {\bibinfo {volume} {8}},\ \bibinfo {pages} {254}
  (\bibinfo {year} {1954})}\BibitemShut {NoStop}%
\bibitem [{\citenamefont {Peel}\ \emph {et~al.}(2012)\citenamefont {Peel},
  \citenamefont {Thompson}, \citenamefont {Daoud-Aladine}, \citenamefont
  {Ashbrook},\ and\ \citenamefont {Lightfoot}}]{peel2012}%
  \BibitemOpen
  \bibfield  {author} {\bibinfo {author} {\bibfnamefont {M.~D.}\ \bibnamefont
  {Peel}}, \bibinfo {author} {\bibfnamefont {S.~P.}\ \bibnamefont {Thompson}},
  \bibinfo {author} {\bibfnamefont {A.}~\bibnamefont {Daoud-Aladine}}, \bibinfo
  {author} {\bibfnamefont {S.~E.}\ \bibnamefont {Ashbrook}}, \ and\ \bibinfo
  {author} {\bibfnamefont {P.}~\bibnamefont {Lightfoot}},\ }\href {\doibase
  10.1021/ic3006585} {\bibfield  {journal} {\bibinfo  {journal} {Inorg. Chem.}\
  }\textbf {\bibinfo {volume} {51}},\ \bibinfo {pages} {6876} (\bibinfo {year}
  {2012})}\BibitemShut {NoStop}%
\bibitem [{\citenamefont {Kennedy}\ \emph {et~al.}(1999)\citenamefont
  {Kennedy}, \citenamefont {Prodjosantoso},\ and\ \citenamefont
  {Howard}}]{kennedy1999}%
  \BibitemOpen
  \bibfield  {author} {\bibinfo {author} {\bibfnamefont {B.~J.}\ \bibnamefont
  {Kennedy}}, \bibinfo {author} {\bibfnamefont {A.}~\bibnamefont
  {Prodjosantoso}}, \ and\ \bibinfo {author} {\bibfnamefont {C.~J.}\
  \bibnamefont {Howard}},\ }\href@noop {} {\bibfield  {journal} {\bibinfo
  {journal} {J. Phys. Cond. Matt.}\ }\textbf {\bibinfo {volume} {11}},\
  \bibinfo {pages} {6319} (\bibinfo {year} {1999})}\BibitemShut {NoStop}%
\bibitem [{\citenamefont {Vogt}\ and\ \citenamefont
  {Schmahl}(1993)}]{vogt1993}%
  \BibitemOpen
  \bibfield  {author} {\bibinfo {author} {\bibfnamefont {T.}~\bibnamefont
  {Vogt}}\ and\ \bibinfo {author} {\bibfnamefont {W.~W.}\ \bibnamefont
  {Schmahl}},\ }\href@noop {} {\bibfield  {journal} {\bibinfo  {journal}
  {Europhys. Lett.}\ }\textbf {\bibinfo {volume} {24}},\ \bibinfo {pages} {281}
  (\bibinfo {year} {1993})}\BibitemShut {NoStop}%
\bibitem [{\citenamefont {Britvin}\ \emph {et~al.}(2022)\citenamefont
  {Britvin}, \citenamefont {Vlasenko}, \citenamefont {Aslandukov},
  \citenamefont {Aslandukov}, \citenamefont {Dubrovinsky}, \citenamefont
  {Gorelova}, \citenamefont {Krzhizhanovskaya}, \citenamefont {Vereshchagin},
  \citenamefont {Bocharov}, \citenamefont {Shelukhina}, \citenamefont
  {Lozhkin}, \citenamefont {Zaitsev},\ and\ \citenamefont
  {Nestola}}]{britvin2022}%
  \BibitemOpen
  \bibfield  {author} {\bibinfo {author} {\bibfnamefont {S.~N.}\ \bibnamefont
  {Britvin}}, \bibinfo {author} {\bibfnamefont {N.~S.}\ \bibnamefont
  {Vlasenko}}, \bibinfo {author} {\bibfnamefont {A.}~\bibnamefont
  {Aslandukov}}, \bibinfo {author} {\bibfnamefont {A.}~\bibnamefont
  {Aslandukov}}, \bibinfo {author} {\bibfnamefont {L.}~\bibnamefont
  {Dubrovinsky}}, \bibinfo {author} {\bibfnamefont {L.~A.}\ \bibnamefont
  {Gorelova}}, \bibinfo {author} {\bibfnamefont {M.~G.}\ \bibnamefont
  {Krzhizhanovskaya}}, \bibinfo {author} {\bibfnamefont {O.~S.}\ \bibnamefont
  {Vereshchagin}}, \bibinfo {author} {\bibfnamefont {V.~N.}\ \bibnamefont
  {Bocharov}}, \bibinfo {author} {\bibfnamefont {Y.~S.}\ \bibnamefont
  {Shelukhina}}, \bibinfo {author} {\bibfnamefont {M.~S.}\ \bibnamefont
  {Lozhkin}}, \bibinfo {author} {\bibfnamefont {A.~N.}\ \bibnamefont
  {Zaitsev}}, \ and\ \bibinfo {author} {\bibfnamefont {F.}~\bibnamefont
  {Nestola}},\ }\href@noop {} {\bibfield  {journal} {\bibinfo  {journal} {Amer.
  Mineral.}\ }\textbf {\bibinfo {volume} {in press}} (\bibinfo {year}
  {2022})}\BibitemShut {NoStop}%
\bibitem [{\citenamefont {Anthony}\ \emph {et~al.}(2005)\citenamefont
  {Anthony}, \citenamefont {Bideaux}, \citenamefont {Bladh},\ and\
  \citenamefont {Nichols}}]{anthony2005}%
  \BibitemOpen
  \bibfield  {author} {\bibinfo {author} {\bibfnamefont {J.~W.}\ \bibnamefont
  {Anthony}}, \bibinfo {author} {\bibfnamefont {R.~A.}\ \bibnamefont
  {Bideaux}}, \bibinfo {author} {\bibfnamefont {K.~W.}\ \bibnamefont {Bladh}},
  \ and\ \bibinfo {author} {\bibfnamefont {M.~C.}\ \bibnamefont {Nichols}},\
  }\href@noop {} {\emph {\bibinfo {title} {Handbook of Mineralogy}}}\ (\bibinfo
   {publisher} {Mineralogical Society of America, Chantilly, VA 20151-1110,
  USA.},\ \bibinfo {year} {2005})\BibitemShut {NoStop}%
\bibitem [{\citenamefont {Yang}\ \emph {et~al.}(1993)\citenamefont {Yang},
  \citenamefont {Ghose},\ and\ \citenamefont {Hatch}}]{yang1993}%
  \BibitemOpen
  \bibfield  {author} {\bibinfo {author} {\bibfnamefont {H.}~\bibnamefont
  {Yang}}, \bibinfo {author} {\bibfnamefont {S.}~\bibnamefont {Ghose}}, \ and\
  \bibinfo {author} {\bibfnamefont {D.}~\bibnamefont {Hatch}},\ }\href@noop {}
  {\bibfield  {journal} {\bibinfo  {journal} {Phys. Chem. Minerals}\ }\textbf
  {\bibinfo {volume} {19}},\ \bibinfo {pages} {528} (\bibinfo {year}
  {1993})}\BibitemShut {NoStop}%
\bibitem [{\citenamefont {Lufaso}\ and\ \citenamefont
  {Woodward}(2001)}]{lufaso2001}%
  \BibitemOpen
  \bibfield  {author} {\bibinfo {author} {\bibfnamefont {M.~W.}\ \bibnamefont
  {Lufaso}}\ and\ \bibinfo {author} {\bibfnamefont {P.~M.}\ \bibnamefont
  {Woodward}},\ }\href@noop {} {\bibfield  {journal} {\bibinfo  {journal} {Acta
  Cryst. B}\ }\textbf {\bibinfo {volume} {57}},\ \bibinfo {pages} {725}
  (\bibinfo {year} {2001})}\BibitemShut {NoStop}%
\bibitem [{\citenamefont {Woodward}(1997{\natexlab{a}})}]{woodward1997a}%
  \BibitemOpen
  \bibfield  {author} {\bibinfo {author} {\bibfnamefont {P.~M.}\ \bibnamefont
  {Woodward}},\ }\href@noop {} {\bibfield  {journal} {\bibinfo  {journal} {Acta
  Cryst. B}\ }\textbf {\bibinfo {volume} {53}},\ \bibinfo {pages} {32}
  (\bibinfo {year} {1997}{\natexlab{a}})}\BibitemShut {NoStop}%
\bibitem [{\citenamefont {Woodward}(1997{\natexlab{b}})}]{woodward1997b}%
  \BibitemOpen
  \bibfield  {author} {\bibinfo {author} {\bibfnamefont {P.~M.}\ \bibnamefont
  {Woodward}},\ }\href@noop {} {\bibfield  {journal} {\bibinfo  {journal} {Acta
  Cryst. B}\ }\textbf {\bibinfo {volume} {53}},\ \bibinfo {pages} {44}
  (\bibinfo {year} {1997}{\natexlab{b}})}\BibitemShut {NoStop}%
\bibitem [{\citenamefont {Yang}\ \emph {et~al.}(2017)\citenamefont {Yang},
  \citenamefont {Wen}, \citenamefont {Xia}, \citenamefont {Sheng},
  \citenamefont {Ma}, \citenamefont {Kim}, \citenamefont {Tapping},
  \citenamefont {Harada}, \citenamefont {Kee}, \citenamefont {Huang} \emph
  {et~al.}}]{yang2017acoustic}%
  \BibitemOpen
  \bibfield  {author} {\bibinfo {author} {\bibfnamefont {J.}~\bibnamefont
  {Yang}}, \bibinfo {author} {\bibfnamefont {X.}~\bibnamefont {Wen}}, \bibinfo
  {author} {\bibfnamefont {H.}~\bibnamefont {Xia}}, \bibinfo {author}
  {\bibfnamefont {R.}~\bibnamefont {Sheng}}, \bibinfo {author} {\bibfnamefont
  {Q.}~\bibnamefont {Ma}}, \bibinfo {author} {\bibfnamefont {J.}~\bibnamefont
  {Kim}}, \bibinfo {author} {\bibfnamefont {P.}~\bibnamefont {Tapping}},
  \bibinfo {author} {\bibfnamefont {T.}~\bibnamefont {Harada}}, \bibinfo
  {author} {\bibfnamefont {T.~W.}\ \bibnamefont {Kee}}, \bibinfo {author}
  {\bibfnamefont {F.}~\bibnamefont {Huang}},  \emph {et~al.},\ }\href@noop {}
  {\bibfield  {journal} {\bibinfo  {journal} {Nature Communications}\ }\textbf
  {\bibinfo {volume} {8}},\ \bibinfo {pages} {1} (\bibinfo {year}
  {2017})}\BibitemShut {NoStop}%
\bibitem [{\citenamefont {Antonius}\ \emph {et~al.}(2015)\citenamefont
  {Antonius}, \citenamefont {Ponc\'e}, \citenamefont {Lantagne-Hurtubise},
  \citenamefont {Auclair}, \citenamefont {Gonze},\ and\ \citenamefont
  {C\^ot\'e}}]{antonius2015}%
  \BibitemOpen
  \bibfield  {author} {\bibinfo {author} {\bibfnamefont {G.}~\bibnamefont
  {Antonius}}, \bibinfo {author} {\bibfnamefont {S.}~\bibnamefont {Ponc\'e}},
  \bibinfo {author} {\bibfnamefont {E.}~\bibnamefont {Lantagne-Hurtubise}},
  \bibinfo {author} {\bibfnamefont {G.}~\bibnamefont {Auclair}}, \bibinfo
  {author} {\bibfnamefont {X.}~\bibnamefont {Gonze}}, \ and\ \bibinfo {author}
  {\bibfnamefont {M.}~\bibnamefont {C\^ot\'e}},\ }\href {\doibase
  10.1103/PhysRevB.92.085137} {\bibfield  {journal} {\bibinfo  {journal} {Phys.
  Rev. B}\ }\textbf {\bibinfo {volume} {92}},\ \bibinfo {pages} {085137}
  (\bibinfo {year} {2015})}\BibitemShut {NoStop}%
\bibitem [{\citenamefont {Adams}\ \emph {et~al.}(2020)\citenamefont {Adams},
  \citenamefont {Wang}, \citenamefont {Steinle-Neumann}, \citenamefont
  {Passerone},\ and\ \citenamefont {Churakov}}]{adams2020}%
  \BibitemOpen
  \bibfield  {author} {\bibinfo {author} {\bibfnamefont {D.~J.}\ \bibnamefont
  {Adams}}, \bibinfo {author} {\bibfnamefont {L.}~\bibnamefont {Wang}},
  \bibinfo {author} {\bibfnamefont {G.}~\bibnamefont {Steinle-Neumann}},
  \bibinfo {author} {\bibfnamefont {D.}~\bibnamefont {Passerone}}, \ and\
  \bibinfo {author} {\bibfnamefont {S.}~\bibnamefont {Churakov}},\ }\href@noop
  {} {\bibfield  {journal} {\bibinfo  {journal} {J. Phys. Cond. Matt.}\
  }\textbf {\bibinfo {volume} {2}},\ \bibinfo {pages} {666} (\bibinfo {year}
  {2020})}\BibitemShut {NoStop}%
\bibitem [{\citenamefont {Adams}\ and\ \citenamefont
  {Oganov}(2006)}]{adams2006}%
  \BibitemOpen
  \bibfield  {author} {\bibinfo {author} {\bibfnamefont {D.~J.}\ \bibnamefont
  {Adams}}\ and\ \bibinfo {author} {\bibfnamefont {A.~R.}\ \bibnamefont
  {Oganov}},\ }\href@noop {} {\bibfield  {journal} {\bibinfo  {journal} {Phys.
  Rev. B}\ }\textbf {\bibinfo {volume} {73}},\ \bibinfo {pages} {184106}
  (\bibinfo {year} {2006})}\BibitemShut {NoStop}%
\bibitem [{\citenamefont {Feynman}(1955)}]{feynman1955}%
  \BibitemOpen
  \bibfield  {author} {\bibinfo {author} {\bibfnamefont {R.~P.}\ \bibnamefont
  {Feynman}},\ }\href@noop {} {\bibfield  {journal} {\bibinfo  {journal} {Phys.
  Rev.}\ }\textbf {\bibinfo {volume} {97}},\ \bibinfo {pages} {660} (\bibinfo
  {year} {1955})}\BibitemShut {NoStop}%
\bibitem [{Note3()}]{Note3}%
  \BibitemOpen
  \bibinfo {note} {This high density is calculated in Ref.~\cite {adams2016}
  for the vibrational DOS of cryolite.}\BibitemShut {Stop}%
\bibitem [{Note4()}]{Note4}%
  \BibitemOpen
  \bibinfo {note} {Data from Bilbao crystallographic server \cite
  {aroyo2006}.}\BibitemShut {Stop}%
\bibitem [{\citenamefont {Aleksandrov}\ and\ \citenamefont
  {Bartolome}(1994)}]{aleksandrov1994}%
  \BibitemOpen
  \bibfield  {author} {\bibinfo {author} {\bibfnamefont {K.~S.}\ \bibnamefont
  {Aleksandrov}}\ and\ \bibinfo {author} {\bibfnamefont {J.}~\bibnamefont
  {Bartolome}},\ }\href {http://stacks.iop.org/0953-8984/6/i=40/a=013}
  {\bibfield  {journal} {\bibinfo  {journal} {J. Phys. Cond. Matt.}\ }\textbf
  {\bibinfo {volume} {6}},\ \bibinfo {pages} {8219} (\bibinfo {year}
  {1994})}\BibitemShut {NoStop}%
\bibitem [{\citenamefont {Steele}\ \emph {et~al.}(2019)\citenamefont {Steele},
  \citenamefont {Puech}, \citenamefont {Monserrat}, \citenamefont {Wu},
  \citenamefont {Yang}, \citenamefont {Kirchartz}, \citenamefont {Yuan},
  \citenamefont {Fleury}, \citenamefont {Giovanni}, \citenamefont {Fron} \emph
  {et~al.}}]{steele2019}%
  \BibitemOpen
  \bibfield  {author} {\bibinfo {author} {\bibfnamefont {J.~A.}\ \bibnamefont
  {Steele}}, \bibinfo {author} {\bibfnamefont {P.}~\bibnamefont {Puech}},
  \bibinfo {author} {\bibfnamefont {B.}~\bibnamefont {Monserrat}}, \bibinfo
  {author} {\bibfnamefont {B.}~\bibnamefont {Wu}}, \bibinfo {author}
  {\bibfnamefont {R.~X.}\ \bibnamefont {Yang}}, \bibinfo {author}
  {\bibfnamefont {T.}~\bibnamefont {Kirchartz}}, \bibinfo {author}
  {\bibfnamefont {H.}~\bibnamefont {Yuan}}, \bibinfo {author} {\bibfnamefont
  {G.}~\bibnamefont {Fleury}}, \bibinfo {author} {\bibfnamefont
  {D.}~\bibnamefont {Giovanni}}, \bibinfo {author} {\bibfnamefont
  {E.}~\bibnamefont {Fron}},  \emph {et~al.},\ }\href@noop {} {\bibfield
  {journal} {\bibinfo  {journal} {ACS Energy Letters}\ }\textbf {\bibinfo
  {volume} {4}},\ \bibinfo {pages} {2205} (\bibinfo {year} {2019})}\BibitemShut
  {NoStop}%
\bibitem [{\citenamefont {Sendner}\ \emph {et~al.}(2016)\citenamefont
  {Sendner}, \citenamefont {Nayak}, \citenamefont {Egger}, \citenamefont
  {Beck}, \citenamefont {M{\"u}ller}, \citenamefont {Epding}, \citenamefont
  {Kowalsky}, \citenamefont {Kronik}, \citenamefont {Snaith}, \citenamefont
  {Pucci} \emph {et~al.}}]{sendner2016}%
  \BibitemOpen
  \bibfield  {author} {\bibinfo {author} {\bibfnamefont {M.}~\bibnamefont
  {Sendner}}, \bibinfo {author} {\bibfnamefont {P.~K.}\ \bibnamefont {Nayak}},
  \bibinfo {author} {\bibfnamefont {D.~A.}\ \bibnamefont {Egger}}, \bibinfo
  {author} {\bibfnamefont {S.}~\bibnamefont {Beck}}, \bibinfo {author}
  {\bibfnamefont {C.}~\bibnamefont {M{\"u}ller}}, \bibinfo {author}
  {\bibfnamefont {B.}~\bibnamefont {Epding}}, \bibinfo {author} {\bibfnamefont
  {W.}~\bibnamefont {Kowalsky}}, \bibinfo {author} {\bibfnamefont
  {L.}~\bibnamefont {Kronik}}, \bibinfo {author} {\bibfnamefont {H.~J.}\
  \bibnamefont {Snaith}}, \bibinfo {author} {\bibfnamefont {A.}~\bibnamefont
  {Pucci}},  \emph {et~al.},\ }\href@noop {} {\bibfield  {journal} {\bibinfo
  {journal} {Materials Horizons}\ }\textbf {\bibinfo {volume} {3}},\ \bibinfo
  {pages} {613} (\bibinfo {year} {2016})}\BibitemShut {NoStop}%
\bibitem [{\citenamefont {Batignani}\ \emph {et~al.}(2018)\citenamefont
  {Batignani}, \citenamefont {Fumero}, \citenamefont {Kandada}, \citenamefont
  {Cerullo}, \citenamefont {Gandini}, \citenamefont {Ferrante}, \citenamefont
  {Petrozza},\ and\ \citenamefont {Scopigno}}]{batignani2018}%
  \BibitemOpen
  \bibfield  {author} {\bibinfo {author} {\bibfnamefont {G.}~\bibnamefont
  {Batignani}}, \bibinfo {author} {\bibfnamefont {G.}~\bibnamefont {Fumero}},
  \bibinfo {author} {\bibfnamefont {A.~R.~S.}\ \bibnamefont {Kandada}},
  \bibinfo {author} {\bibfnamefont {G.}~\bibnamefont {Cerullo}}, \bibinfo
  {author} {\bibfnamefont {M.}~\bibnamefont {Gandini}}, \bibinfo {author}
  {\bibfnamefont {C.}~\bibnamefont {Ferrante}}, \bibinfo {author}
  {\bibfnamefont {A.}~\bibnamefont {Petrozza}}, \ and\ \bibinfo {author}
  {\bibfnamefont {T.}~\bibnamefont {Scopigno}},\ }\href@noop {} {\bibfield
  {journal} {\bibinfo  {journal} {Nature communications}\ }\textbf {\bibinfo
  {volume} {9}},\ \bibinfo {pages} {1} (\bibinfo {year} {2018})}\BibitemShut
  {NoStop}%
\bibitem [{\citenamefont {Prodanovi{\'c}}\ and\ \citenamefont
  {Vukmirovi{\'c}}(2019)}]{prodanovic2019}%
  \BibitemOpen
  \bibfield  {author} {\bibinfo {author} {\bibfnamefont {N.}~\bibnamefont
  {Prodanovi{\'c}}}\ and\ \bibinfo {author} {\bibfnamefont {N.}~\bibnamefont
  {Vukmirovi{\'c}}},\ }\href@noop {} {\bibfield  {journal} {\bibinfo  {journal}
  {Phys. Rev. B}\ }\textbf {\bibinfo {volume} {99}},\ \bibinfo {pages} {104304}
  (\bibinfo {year} {2019})}\BibitemShut {NoStop}%
\bibitem [{\citenamefont {Herz}(2017)}]{herz2017}%
  \BibitemOpen
  \bibfield  {author} {\bibinfo {author} {\bibfnamefont {L.~M.}\ \bibnamefont
  {Herz}},\ }\href@noop {} {\bibfield  {journal} {\bibinfo  {journal} {ACS
  Energy Lett.}\ }\textbf {\bibinfo {volume} {2}},\ \bibinfo {pages} {1539}
  (\bibinfo {year} {2017})}\BibitemShut {NoStop}%
\bibitem [{\citenamefont {Fr{\"o}hlich}(1954)}]{frohlich1954}%
  \BibitemOpen
  \bibfield  {author} {\bibinfo {author} {\bibfnamefont {H.}~\bibnamefont
  {Fr{\"o}hlich}},\ }\href@noop {} {\bibfield  {journal} {\bibinfo  {journal}
  {Advances in Physics}\ }\textbf {\bibinfo {volume} {3}},\ \bibinfo {pages}
  {325} (\bibinfo {year} {1954})}\BibitemShut {NoStop}%
\bibitem [{\citenamefont {Iaru}\ \emph {et~al.}(2017)\citenamefont {Iaru},
  \citenamefont {Geuchies}, \citenamefont {Koenraad}, \citenamefont
  {Vanmaekelbergh},\ and\ \citenamefont {Silov}}]{iaru2017}%
  \BibitemOpen
  \bibfield  {author} {\bibinfo {author} {\bibfnamefont {C.~M.}\ \bibnamefont
  {Iaru}}, \bibinfo {author} {\bibfnamefont {J.~J.}\ \bibnamefont {Geuchies}},
  \bibinfo {author} {\bibfnamefont {P.~M.}\ \bibnamefont {Koenraad}}, \bibinfo
  {author} {\bibfnamefont {D.}~\bibnamefont {Vanmaekelbergh}}, \ and\ \bibinfo
  {author} {\bibfnamefont {A.~Y.}\ \bibnamefont {Silov}},\ }\href@noop {}
  {\bibfield  {journal} {\bibinfo  {journal} {ACS nano}\ }\textbf {\bibinfo
  {volume} {11}},\ \bibinfo {pages} {11024} (\bibinfo {year}
  {2017})}\BibitemShut {NoStop}%
\bibitem [{\citenamefont {Kirschner}\ \emph {et~al.}(2019)\citenamefont
  {Kirschner}, \citenamefont {Diroll}, \citenamefont {Guo}, \citenamefont
  {Harvey}, \citenamefont {Helweh}, \citenamefont {Flanders}, \citenamefont
  {Brumberg}, \citenamefont {Watkins}, \citenamefont {Leonard}, \citenamefont
  {Evans} \emph {et~al.}}]{kirschner2019}%
  \BibitemOpen
  \bibfield  {author} {\bibinfo {author} {\bibfnamefont {M.~S.}\ \bibnamefont
  {Kirschner}}, \bibinfo {author} {\bibfnamefont {B.~T.}\ \bibnamefont
  {Diroll}}, \bibinfo {author} {\bibfnamefont {P.}~\bibnamefont {Guo}},
  \bibinfo {author} {\bibfnamefont {S.~M.}\ \bibnamefont {Harvey}}, \bibinfo
  {author} {\bibfnamefont {W.}~\bibnamefont {Helweh}}, \bibinfo {author}
  {\bibfnamefont {N.~C.}\ \bibnamefont {Flanders}}, \bibinfo {author}
  {\bibfnamefont {A.}~\bibnamefont {Brumberg}}, \bibinfo {author}
  {\bibfnamefont {N.~E.}\ \bibnamefont {Watkins}}, \bibinfo {author}
  {\bibfnamefont {A.~A.}\ \bibnamefont {Leonard}}, \bibinfo {author}
  {\bibfnamefont {A.~M.}\ \bibnamefont {Evans}},  \emph {et~al.},\ }\href@noop
  {} {\bibfield  {journal} {\bibinfo  {journal} {Nature communications}\
  }\textbf {\bibinfo {volume} {10}},\ \bibinfo {pages} {1} (\bibinfo {year}
  {2019})}\BibitemShut {NoStop}%
\bibitem [{\citenamefont {Glazer}\ \emph {et~al.}(1993)\citenamefont {Glazer},
  \citenamefont {Roleder},\ and\ \citenamefont {Dec}}]{glazer1993}%
  \BibitemOpen
  \bibfield  {author} {\bibinfo {author} {\bibfnamefont {A.}~\bibnamefont
  {Glazer}}, \bibinfo {author} {\bibfnamefont {K.}~\bibnamefont {Roleder}}, \
  and\ \bibinfo {author} {\bibfnamefont {J.}~\bibnamefont {Dec}},\ }\href@noop
  {} {\bibfield  {journal} {\bibinfo  {journal} {Acta Cryst. B}\ }\textbf
  {\bibinfo {volume} {49}},\ \bibinfo {pages} {846} (\bibinfo {year}
  {1993})}\BibitemShut {NoStop}%
\bibitem [{\citenamefont {Aroyo}\ \emph {et~al.}(2006)\citenamefont {Aroyo},
  \citenamefont {Perez-Mato}, \citenamefont {Capillas}, \citenamefont
  {Kroumova}, \citenamefont {Ivantchev}, \citenamefont {Madariaga},
  \citenamefont {Kirov},\ and\ \citenamefont {Wondratschek}}]{aroyo2006}%
  \BibitemOpen
  \bibfield  {author} {\bibinfo {author} {\bibfnamefont {M.~I.}\ \bibnamefont
  {Aroyo}}, \bibinfo {author} {\bibfnamefont {J.~M.}\ \bibnamefont
  {Perez-Mato}}, \bibinfo {author} {\bibfnamefont {C.}~\bibnamefont
  {Capillas}}, \bibinfo {author} {\bibfnamefont {E.}~\bibnamefont {Kroumova}},
  \bibinfo {author} {\bibfnamefont {S.}~\bibnamefont {Ivantchev}}, \bibinfo
  {author} {\bibfnamefont {G.}~\bibnamefont {Madariaga}}, \bibinfo {author}
  {\bibfnamefont {A.}~\bibnamefont {Kirov}}, \ and\ \bibinfo {author}
  {\bibfnamefont {H.}~\bibnamefont {Wondratschek}},\ }\href@noop {} {\bibfield
  {journal} {\bibinfo  {journal} {Z. Kristallographie}\ }\textbf {\bibinfo
  {volume} {221}},\ \bibinfo {pages} {15} (\bibinfo {year} {2006})}\BibitemShut
  {NoStop}%
\end{thebibliography}%
% \bibstyle{iucr} %\bibliographystyle{iucr}  %\bibliographystyle{plainnat}
 %\bibliography{/home/adj127/literature/bibliography}

% The following MDPI journals use author-date citation: Arts, Econometrics, Economies, Genealogy, Humanities, IJFS, JRFM, Laws, Religions, Risks, Social Sciences. For those journals, please follow the formatting guidelines on http://www.mdpi.com/authors/references
% To cite two works by the same author: \citeauthor{ref-journal-1a} (\citeyear{ref-journal-1a}, \citeyear{ref-journal-1b}). This produces: Whittaker (1967, 1975)
% To cite two works by the same author with specific pages: \citeauthor{ref-journal-3a} (\citeyear{ref-journal-3a}, p. 328; \citeyear{ref-journal-3b}, p.475). This produces: Wong (1999, p. 328; 2000, p. 475)

%% for journal Sci
%\reviewreports{\\
%Reviewer 1 comments and authors' response\\
%Reviewer 2 comments and authors' response\\
%Reviewer 3 comments and authors' response
%}

%%%%%%%%%%%%%%%%%%%%%%%%%%%%%%%%%%%%%%%%%%
\end{document}